\documentclass[prd,preprintnumbers,amsmath,amssymb,nofootinbib]{revtex4}
\usepackage{graphics,color,subfigure}
\usepackage{epsfig}
\usepackage{dcolumn}
\usepackage{bm}
\usepackage{longtable}
\usepackage{multirow}
\usepackage{diagbox}
\usepackage{amsmath}
\usepackage[colorlinks=true,
urlcolor=blue,
linkcolor=blue,
citecolor=blue,
bookmarks=true,
bookmarksopen=true,
bookmarksnumbered=true]{hyperref}

\begin{document}
	\title{Triply heavy tetraquark states in a mass-splitting model}
	\author{Shi-Yuan Li$^1$}
	\author{Yan-Rui Liu$^1$}\email{yrliu@sdu.edu.cn}
	\author{Zi-Long Man$^{2,3,4,5,6}$}\email{manzl@lzu.edu.cn}
	\author{Cheng-Rui Shu$^1$}
	\author{Zong-Guo Si$^1$}
	\author{Jing Wu$^7$}\email{wujing18@sdjzu.edu.cn}
	\affiliation{$^1$School of Physics, Shandong University, Jinan, Shandong 250100, China\\
		$^2$School of Physical Science and Technology, Lanzhou University, Lanzhou 730000, China\\
		$^3$Lanzhou Center for Theoretical Physics, Lanzhou University, Lanzhou 730000, China\\
		$^4$Key Laboratory of Quantum Theory and Applications of MoE, Lanzhou University, Lanzhou 730000, China\\
		$^5$Key Laboratory of Theoretical Physics of Gansu Province, Lanzhou University, Lanzhou 730000, China\\
		$^6$Research Center for Hadron and CSR Physics, Lanzhou University and Institute of Modern Physics of CAS, Lanzhou 730000, China\\
		$^7$School of Science, Shandong Jianzhu University, Jinan 250101, China
	}
	\date{\today}
	
\begin{abstract}
In a modified chromomagnetic interaction model, assuming $X(4140)$ to be the lowest $1^{++}$ $cs\bar{c}\bar{s}$ tetraquark and treating it as the reference state, we systematically investigated the masses of the triply-heavy tetraquark states $QQ\bar{Q}\bar{q}$ ($Q=c,b;q=u,d,s$). Because of their higher masses, no stable tetraquarks were found. Using a simple scheme, we also estimated the partial widths of the rearrangement decay channels and relevant ratios. A compact triply heavy tetraquark candidate would be favored if its observed mass and partial width ratios were comparable with our predictions. We hope that the present work will be helpful for further studies.
\end{abstract}
	
\maketitle

\section{Introduction}\label{secI}
	
Since the observation of the exotic $X(3872)$ by the Belle Collaboration in 2003 \cite{Belle:2003nnu}, a large number of charmonium-like and bottomonium-like states with the names $X$, $Y$, or $Z$ have been identified over the past two decades \cite{Belle:2005lik,Belle:2007woe,Belle:2009and,Belle:2009rkh,CDF:2009jgo,LHCb:2016axx,LHCb:2016nsl,BESIII:2016bnd,BESIII:2016adj,Belle:2017egg,LHCb:2018oeg}. In particular, charged charmonium-like or bottomonium-like states were found, such as $Z_c(3900)$ \cite{BESIII:2013ris,Belle:2013yex,Xiao:2013iha,BESIII:2015cld}, $Z_c(3885)$ \cite{BESIII:2013qmu,BESIII:2015pqw}, $Z_c(4020)$ \cite{BESIII:2013ouc,BESIII:2014gnk}, $Z_c(4025)$ \cite{BESIII:2013mhi,BESIII:2015tix}, $Z_{cs}(3985)$ \cite{BESIII:2020qkh}, $Z_{cs}(4000)$ \cite{LHCb:2021uow}, $Z_{cs}(4220)$ \cite{LHCb:2021uow}, $Z_b(10610)$ \cite{Belle:2011aa}, and $Z_b(10650)$ \cite{Belle:2011aa}, which cannot be classified as excited heavy quarkonia and are explicitly exotic. Their properties may be understood in  configurations such as the compact tetraquark \cite{Maiani:2004vq,Ebert:2007rn,Anwar:2018sol}, the meson-antimeson molecule \cite{Tornqvist:1993ng,Tornqvist:2004qy,Swanson:2003tb,Hanhart:2007yq}. In 2020, the LHCb Collaboration observed a broad structure that ranged from 6.2 to 6.8 GeV and a narrow one located around 6.9 GeV in the $J/\psi J/\psi$ channel, while the latter was called $X(6900)$ \cite{LHCb:2020bwg}. They are good candidates for fully heavy tetraquark states. More candidates were announced in Refs. \cite{CMS:2023owd,ATLAS:2023bft,ANDY:2019bfn}. It is essential to study the exotic structures from broader and deeper perspectives  \cite{Chen:2016qju,Esposito:2016noz,Lebed:2016hpi,Ali:2017jda,Olsen:2017bmm,Guo:2017jvc,Yuan:2018inv,Brambilla:2019esw,Liu:2019zoy,Chen:2022asf,Liu:2024uxn}. 

In addition to the hidden heavy case, open heavy exotics have been also observed in recent years. In 2016, the D0 Collaboration reported the observation of the singly-bottom $X(5568)$ in the $B_s^0\pi^{\pm}$ channel \cite{D0:2016mwd}. This four-quark state is about 200 MeV below the $B\bar{K}$ threshold. However, the LHCb Collaboration \cite{LHCb:2016dxl} and the CMS Collaboration did not corroborate the presence of this state, casting doubt on its existence. In 2020, the LHCb observed an exotic peak in the $D^-K^+$ channel \cite{LHCb:2020bls,LHCb:2020pxc}. To fit the experimental data, the collaboration introduced two resonances, named $T_{cs0}(2900)^0$ $(J=0)$ and $T_{cs1}(2900)^0$ $(J=1)$, whose minimal quark content is $ud\bar{s}\bar{c}$. In 2023, they observed another two singly charmed tetraquark states, $T_{c\bar{s}0}^a(2900)^0$ and $T_{c\bar{s}0}^a(2900)^{++}$ \cite{LHCb:2022sfr,LHCb:2022lzp}. In the double-charm case, the LHCb also produced a new finding. In 2021, they reported the observation of a narrow state named $T_{cc}^+(3875)$ in the $D^0D^0\pi^+$ mass spectrum, just below the $D^{*+}D^0$ threshold \cite{LHCb:2021vvq,LHCb:2021auc}. This state has a minimal quark content of $cc\bar{u}\bar{d}$ and is a good candidate for the theoretically anticipated double-charm tetraquark $T_{cc}$ \cite{Liu:2019zoy}. 
	
Until now, possible singly, doubly, and fully charm tetraquark states have been observed. The existence of triply heavy tetraquarks is also possible. Distinguishing between compact states and hadronic molecules is often challenging for researchers. However, in the case of the fully heavy state $QQ\bar{Q}\bar{Q}$ ($Q=c,b$), the meson-exchange interaction may be suppressed, while the short-range one-gluon-exchange interaction should play a dominant role in the binding force. It is very likely that the observed $X(6900)$ is a compact tetraquark. The situation is similar for the triply heavy $QQ\bar{Q}\bar{q}$ ($q=u,d,s$) states. If such a state were observed in future experiments, understanding its properties in a compact picture is highly feasible. The authors of Refs. \cite{Liu:2023gla,Liu:2024pio} considered the possibility of fully heavy four-quark molecular states using heavy meson exchange forces. If such interactions do play an important role, triply heavy four-quark molecular states should also be possible.
	
To date, there have been several theoretical explorations of triply heavy tetraquark states using various methods. With the assumption that the input $X(3872)$ is a tetraquark state, triply heavy tetraquark spectra were studied in Ref. \cite{Cui:2006mp} using a chromomagnetic interaction (CMI) model and some stable states were found. A different CMI model adopted in Ref. \cite{Chen:2016ont} also gave some stable states. However, unstable states were obtained with an extended CMI model in Ref. \cite{Weng:2021ngd}. From calculations utilizing lattice QCD \cite{Junnarkar:2018twb,Hudspith:2020tdf}, shallow bound $uc\bar{b}\bar{b}$ and $sc\bar{b}\bar{b}$ states are possible. A study that used the QCD sum rule \cite{Jiang:2017tdc} indicated that narrow resonances are possible, while a recent calculation \cite{Zhang:2024jvv} gave heavier $cc\bar{c}\bar{q}/bb\bar{b}\bar{q}$ ($q=u,d,s$) states. Nonstrange multiquarks, as compact topological molecules, were studied using a holographic approach in Ref. \cite{Liu:2019mxw}, which revealed that the $QQ\bar{Q}\bar{q}$ states are unbound. Stable candidates were obtained in the $cc\bar{c}\bar{n}$ sector using an AdS/QCD potential model \cite{Mutuk:2023yev}. Using the MIT bag model, Ref. \cite{Zhu:2023lbx} indicated that all the triply heavy tetraquarks are above the corresponding meson-meson thresholds. The authors of Ref. \cite{Lu:2021kut} also drew the conclusion that there are no stable $QQ\bar{Q}\bar{q}$ states by employing an extended relativized quark model. A similar conclusion was obtained in Ref. \cite{Meng:2023jqk}, where a pure constituent quark model and a chiral constituent quark model were employed. With their constituent quark models, the authors of Refs. \cite{Liu:2022jdl,Yang:2024nyc} found that bound $cc\bar{c}\bar{n}$ states are possible when coupled channel effects are considered. In addition to mass calculations, the authors of Ref. \cite{Xing:2019wil} employed two models, namely, the effective Hamiltonian in the diquark-antidiquark picture and the nonrelativistic quark model, to study the decay properties of $bb\bar{c}\bar{n}$ ($n=u,d$) states.

When determining the spectra of triply heavy tetraquark states with a CMI model in Ref. \cite{Chen:2016ont}, we utilized meson-meson thresholds as reference scales. Several states below the corresponding lowest meson-meson thresholds were found, which indicates that they may be stable; e.g., the lowest $1^+$ $cc\bar{b}\bar{n}$ was below the $B_cD^*$ threshold. However, the obtained tetraquark masses may have been underestimated \cite{Chen:2016ont,Cheng:2020nho,Wu:2018xdi}. To reduce the estimation uncertainty, following the study method for tetraquarks adopted in Refs. \cite{Cheng:2020nho,Li:2023wxm,Wu:2018xdi,Li:2023wug}, we reconsidered the $QQ\bar{Q}\bar{q}$ ($q=u,d,s$) spectra by treating the $1^{++}$ $X(4140)$ \cite{CDF:2009jgo,LHCb:2016axx} as the reference tetraquark state. Since the inner structures between the meson-meson and compact states are different, the masses estimated using this method should be more reasonable. We also discuss the two-body rearrangement decays, which were not considered in Ref. \cite{Chen:2016ont}, by employing a simple scheme.

This paper is organized as follows: In Sec. \ref{secII}, we present the formalism containing the mass formulae, color-spin base vectors, CMI matrices for different systems, and scheme to study the rearrangement decays. In Sec. \ref{secIII}, we collect the parameters for the calculation and the numerical results, including the spectra and rearrangement decay widths. We provide a discussion and a short summary in the last section.

\section{Formalism}\label{secII}

\subsection{Spectrum calculation}\label{massformu}
In this work, we employed the CMI model to study the S-wave triply heavy tetraquark states. The model Hamiltonian reads as
\begin{eqnarray}\label{hamiltonian}
H=\sum_i m_i +H_{CMI}=\sum_i m_i -\sum_{i<j} C_{ij}\lambda_i\cdot\lambda_{j}\sigma_{i}\cdot\sigma_{j}.
\end{eqnarray}
Here, $m_i$ is the effective mass of the $i$th quark component, which contains contributions from the kinetic energy, color--Coulomb potential, and color confinement. The effective parameter $C_{ij}$ reflects the coupling strength between the $i$th and $j$th quark components. $\lambda_{i}$ and $\sigma_{i}$ are the Gell--Mann and Pauli matrices, respectively, for the $i$th quark. For antiquarks, $\lambda_i$ should be replaced with $-\lambda_{i}^{*}$. The chromomagnetic term $H_{CMI}$ induces mass splittings for the tetraquark states. With the constructed color--spin base vectors, the CMI matrix $\langle H_{CMI}\rangle$ can be obtained; diagonalizing it gives the mass formula for a compact tetraquark:
\begin{eqnarray}\label{Mass}
M=\sum_i m_i +E_{CMI},
\end{eqnarray}
where $E_{CMI}$ indicates the eigenvalue of $\langle{H_{CMI}}\rangle$ corresponding to this state.
	
Since the effective quark masses are extracted from the spectra of conventional mesons and baryons, they may not be suitable for tetraquark states. From previous studies \cite{Liu:2019zoy,Wu:2018xdi,Cheng:2020nho,Li:2023wxm,Li:2023wug}, we found that the values calculated using Eq. \eqref{Mass} tend to be larger than the possible tetraquark masses. This discrepancy is primarily attributed to the values of effective quark masses, which may not accurately reflect the interactions within tetraquark states. Each hadron has quark masses tailored to its specific structure, and the extracted values may not be directly applicable to multiquark systems. The overestimated tetraquark masses from Eq. \eqref{Mass} are regarded as the theoretical upper limits in the following discussions. 

To reduce the uncertainties and obtain more reasonable tetraquark spectra, one may adopt a modified mass formula by introducing a reference state that has the same quark content as the studied tetraquark,
\begin{eqnarray}\label{massref}
M=[M_{ref}-(E_{CMI})_{ref}]+E_{CMI}.
\end{eqnarray}
Here, $M_{ref}$ and $(E_{CMI})_{ref}$ denote the measured mass and the calculated CMI eigenvalue for the reference state, respectively. One of the choices for the reference scale $M_{ref}$ for a considered system is a meson-meson threshold. However, previous studies \cite{Chen:2016ont,Cheng:2020nho,Wu:2018xdi} indicated that such a choice is not unique and may result in tetraquark masses lower than the measured values. The reason should be that interactions between constituent quarks in compact multiquark states are complex and cannot be fully reflected in a simple hadron-hadron state. We regard the underestimated tetraquark masses from Eq. \eqref{massref} as the theoretical lower limits in the following.

To obtain more reasonable values, it is necessary to choose a reference scale for all tetraquark mass estimations. Considering that the dynamics of two tetraquark states are comparable, it is reasonable to select a tetraquark candidate to determine the scale. In previous studies \cite{Cheng:2020nho,Li:2023wxm,Wu:2018xdi,Li:2023wug}, we treated $X(4140)$ as the reference by assuming it to be the lowest $1^{++}$ compact $cs\bar{c}\bar{s}$ state. In this work, we again adopted this assumption. The considerations were as follows. First, $X(4140)$ as a $J/\psi\phi$ resonance was confirmed by different experiments with the determined quantum numbers $J^{PC}=1^{++}$. Secondly, the exotic state $X(4274)$ was observed in the $J/\psi\phi$ channel by CDF and LHCb \cite{LHCb:2021uow,CDF:2011pep} with the same quantum numbers as $X(4140)$. These states can be interpreted consistently as partner states in the compact $cs\bar{c}\bar{s}$ picture \cite{Wu:2016gas,Stancu:2009ka}. Additionally, from discussions on the reference selection problem in Ref. \cite{Li:2023wxm}, we found that adopting $X(4140)$ as the reference can give more reasonable interpretations for other $cs\bar{c}\bar{s}$ states. Now, a second modified mass formula reads as
\begin{equation}\label{mass4140}
\begin{split}
M&=M_{X(4140)}-(E_{CMI})_{X(4140)}+E_{CMI} +\sum_{ij}n_{ij}(m_i-m_j)\\
&=\tilde{m}+E_{CMI}+\sum_{ij}n_{ij}\Delta_{ij}.
\end{split}
\end{equation} 
Here, $M_{X(4140)}$ and $(E_{CMI})_{X(4140)}$ are the measured mass and calculated CMI eigenvalue of $X(4140)$, respectively. The quark contents are different for triply heavy tetraquarks and the hidden-charm $X(4140)$. This modified formula means that we used the quark mass gap $\Delta_{ij}=m_i-m_j$, rather than the quark masses themselves, as well as the integer number $n_{ij}$ to parameterize the scale difference. The value of $\Delta_{ij}$ will be extracted from the conventional hadron masses. Explicitly, the mass formulae for the systems considered in this study are
\begin{eqnarray}\label{mass4140detail}
\begin{array}{ll}
M_{cc\bar{c}\bar{n}}=\tilde{m}+\langle{H_{CMI}}\rangle+\Delta_{cs}-\Delta_{sn},\\
M_{cc\bar{c}\bar{s}}=\tilde{m}+\langle{H_{CMI}}\rangle+\Delta_{cs},\\
M_{cc\bar{b}\bar{n}}=\tilde{m}+\langle{H_{CMI}}\rangle+\Delta_{bs}-\Delta_{sn},\\
M_{cc\bar{b}\bar{s}}=\tilde{m}+\langle{H_{CMI}}\rangle+\Delta_{bs},\\
M_{bb\bar{c}\bar{n}}=\tilde{m}+\langle{H_{CMI}}\rangle+2\Delta_{bs}-\Delta_{cn},\\
M_{bb\bar{c}\bar{s}}=\tilde{m}+\langle{H_{CMI}}\rangle+\Delta_{bc}+\Delta_{bs},\\
M_{bb\bar{b}\bar{n}}=\tilde{m}+\langle{H_{CMI}}\rangle+2\Delta_{bs}+\Delta_{bc}-\Delta_{cn},\\
M_{bb\bar{b}\bar{s}}=\tilde{m}+\langle{H_{CMI}}\rangle+2\Delta_{bc}+\Delta_{bs},\\
M_{bc\bar{c}\bar{n}}=M_{cc\bar{b}\bar{n}},\quad 
M_{bc\bar{c}\bar{s}}=M_{cc\bar{b}\bar{s}},\\
M_{bc\bar{b}\bar{n}}=M_{bb\bar{c}\bar{n}},\quad
M_{bc\bar{b}\bar{s}}=M_{bb\bar{c}\bar{s}}.
\end{array}
\end{eqnarray}
Although the formulae for different systems may have been the same, the number of states and the mass spectra were not. We discuss the results calculated with these formulae.

\subsection{Color-spin base vectors and CMI Hamiltonians}\label{basevector}
	
It is essential to establish color$\otimes$spin base vectors to obtain the CMI matrices. We chose the diquark-antidiquark configuration to describe the bases. They are the same as those in Ref. \cite{Chen:2016ont},
\begin{eqnarray}
\begin{array}{ll}
\phi_1\chi_1=|(Q_1Q_2)_1^6(\bar{Q}_3\bar{q}_4)_1^{\bar{6}}\rangle_2\delta_{12},\\
\phi_1\chi_2=|(Q_1Q_2)_1^6(\bar{Q}_3\bar{q}_4)_1^{\bar{6}}\rangle_1\delta_{12},\\
\phi_1\chi_3=|(Q_1Q_2)_1^6(\bar{Q}_3\bar{q}_4)_1^{\bar{6}}\rangle_0\delta_{12},\\
\phi_1\chi_4=|(Q_1Q_2)_1^6(\bar{Q}_3\bar{q}_4)_0^{\bar{6}}\rangle_1\delta_{12},\\
\phi_1\chi_5=|(Q_1Q_2)_0^6(\bar{Q}_3\bar{q}_4)_1^{\bar{6}}\rangle_1,\\
\phi_1\chi_6=|(Q_1Q_2)_0^6(\bar{Q}_3\bar{q}_4)_0^{\bar{6}}\rangle_0,\\
\phi_2\chi_1=|(Q_1Q_2)_1^{\bar{3}}(\bar{Q}_3\bar{q}_4)_1^3\rangle_2,\\
\phi_2\chi_2=|(Q_1Q_2)_1^{\bar{3}}(\bar{Q}_3\bar{q}_4)_1^3\rangle_1,\\
\phi_2\chi_3=|(Q_1Q_2)_1^{\bar{3}}(\bar{Q}_3\bar{q}_4)_1^3\rangle_0,\\
\phi_2\chi_4=|(Q_1Q_2)_1^{\bar{3}}(\bar{Q}_3\bar{q}_4)_0^3\rangle_1,\\
\phi_2\chi_5=|(Q_1Q_2)_0^{\bar{3}}(\bar{Q}_3\bar{q}_4)_1^3\rangle_1\delta_{12},\\
\phi_2\chi_6=|(Q_1Q_2)_0^{\bar{3}}(\bar{Q}_3\bar{q}_4)_0^3\rangle_0\delta_{12},
\end{array}
\end{eqnarray}
where $\phi$ and $\chi$ are color and spin base vectors, respectively, and the notation on the right-hand side is $|(Q_1Q_2)_{spin}^{color}(\bar{Q_3}\bar{q_4})_{spin}^{color}\rangle_{spin}$. The $\delta_{12}$ symbol arises from the Pauli principle. It is set to 0 if $Q_1$ and $Q_2$ are identical, otherwise, it is equal to 1. This convention means that the corresponding base vector does not exist for states with $Q_1=Q_2$. Therefore, we could categorize the studied systems into two groups: one contains $cc\bar{Q}\bar{q}$ and $bb\bar{Q}\bar{q}$ systems and the other contains $bc\bar{Q}\bar{q}$. The first group involves six base vectors, but all the twelve bases are involved in the second group.

To express the matrices succinctly, here we define $\alpha=C_{12}+C_{34}$, $\gamma=C_{12}-C_{34}$, $\beta=C_{13}+C_{14}+C_{23}+C_{24}$, $\delta=C_{13}-C_{14}+C_{23}-C_{24}$, $\mu=C_{13}-C_{14}-C_{23}+C_{24}$, and $\nu=C_{13}+C_{14}-C_{23}-C_{24}$. For the $2^+$, $1^+$, and $0^+$ states in the first group, the corresponding CMI matrices are $\frac43(2\alpha+\beta)$ with the base vector $(\phi_2\chi_1)^T$,
\begin{eqnarray}
\left(\begin{array}{ccc}
\frac43(2\alpha-\beta)&\frac43\sqrt{2}\delta&4\delta\\&\frac83(2\gamma-\alpha)&-2\sqrt{2}\beta\\&&\frac43(\alpha+2\gamma)
\end{array}\right)
\end{eqnarray}
with the base vector $(\phi_2\chi_2, \phi_2\chi_4, \phi_1\chi_5)^T$, and
\begin{eqnarray}
\left(\begin{array}{cc}
\frac83(\alpha-\beta)&2\sqrt{6}\beta\\&4\alpha
\end{array}\right)
\end{eqnarray}
with the base vector $(\phi_2\chi_3, \phi_1\chi_6)^T$. For the $2^+$, $1^+$, and $0^+$ states in the second group, the corresponding CMI matrices are
\begin{eqnarray}
\left(\begin{array}{cc}
-\frac43\alpha+\frac{10}{3}\beta&-2\sqrt{2}\mu\\&\frac43(2\alpha+\beta)
\end{array}\right),
\end{eqnarray}
\begin{eqnarray}
\left(\begin{array}{cccccc}
-\frac43\alpha-\frac{10}{3}\beta&\frac{10}{3}\sqrt{2}\delta&-\frac{10}{3}\sqrt{2}\nu&2\sqrt{2}\mu&-4\nu&4\delta\\&\frac43(\alpha-2\gamma)&\frac{10}{3}\mu&-4\nu&0&-2\sqrt{2}\beta\\&&\frac43(\alpha+2\gamma)&4\delta&-2\sqrt{2}\beta&0\\&&&\frac43(2\alpha-\beta)&\frac43\sqrt{2}\delta&-\frac43\sqrt{2}\nu\\&&&&\frac83(2\gamma-\alpha)&\frac43\mu\\&&&&&-\frac83(\alpha+2\gamma)
\end{array}\right),
\end{eqnarray}
and
\begin{eqnarray}
\left(\begin{array}{cccc}
\frac83(\alpha-\beta)&-\frac43\sqrt{3}\mu&4\sqrt{2}\mu&2\sqrt{6}\beta\\&-8\alpha&2\sqrt{6}\beta&0\\&&-\frac43(\alpha+5\beta)&-\frac{10}{3}\sqrt{3}\mu\\&&&4\alpha
\end{array}\right).
\end{eqnarray}
Their base vectors are $(\phi_1\chi_1, \phi_2\chi_1)^T$, $(\phi_1\chi_2, \phi_1\chi_4, \phi_1\chi_5, \phi_2\chi_2, \phi_2\chi_4, \phi_2\chi_5)^T$, and $(\phi_2\chi_3, \phi_2\chi_6, \phi_1\chi_3, \phi_1\chi_6)^T$, respectively. Since each CMI matrix is symmetric, here we only write down the upper triangular part.

\subsection{Rearrangement decay}

Our study also involves the rearrangement decays, which were found to be helpful in understanding exotic hadron structures by combining information from spectra and decay widths \cite{Cheng:2019obk,Cheng:2020nho,Li:2023wxm,Li:2023aui}. The simple scheme that we adopt is just to estimate the scattering amplitude $\mathcal{M}$ by taking the decay Hamiltonian as a constant $H_{decay}={\cal C}$. Then, the amplitude is written as $\mathcal{M}=\langle{final}|H_{decay}|initial\rangle={\cal C}\langle{final}|initial\rangle$ and the decay width is
\begin{eqnarray}\label{decayformula}
\Gamma=|\mathcal{M}|^2\frac{|\vec{p}|}{8\pi M^2},
\end{eqnarray}
where $M$ is the initial tetraquark state mass and $\vec{p}$ is the three-momentum of a final meson in the center-of-mass frame. To obtain the $\mathcal{M}$ values, we need the flavor-color-spin wave functions of $|initial\rangle$ and $|final\rangle$. The final state has two possible configurations: $(Q_1\bar{Q}_3)^{1c}(Q_2\bar{q}_4)^{1c}$ and $(Q_1\bar{q}_4)^{1c}(Q_2\bar{Q}_3)^{1c}$. They could be expressed as superpositions of the base vectors given in the last subsection with different coefficients. Supposing that $|initial\rangle=\sum_{i=1}^{12}x_i\psi_i$ and $|final\rangle=\sum_{i=1}^{12}y_i\psi_i$, where $\psi_i$ is the $i$th base vector, we obtain $\mathcal{M}={\cal C} \sum_{i=1}^{12}x_iy_i$ and $\Gamma$ immediately.

\section{Spectra and widths of triply heavy tetraquarks}\label{secIII}

\subsection{Model parameters}

\begin{table}[htbp]\centering
\renewcommand{\arraystretch}{1.2}
\caption{Effective coupling parameters $C_{ij}$ in MeV units.}\label{coupling}
\setlength{\tabcolsep}{5mm}{
		\begin{tabular}{cccccc}\hline\hline
			$C_{ij}$ & $c$ & $b$ & $C_{i\bar{j}}$ & $\bar{c}$ & $\bar{b}$ \\\hline
			$n$ & 4.0 & 1.3 & $n$ & 6.6 & 2.1 \\
			$s$ & 4.3 & 1.3 & $s$ & 6.7 & 2.3 \\
			$c$ & 3.2 & 2.0 & $c$ & 5.3 & 3.3 \\
			$b$ &  & 1.9 & $b$ &  & 2.9 \\\hline\hline
	\end{tabular}}
\end{table}

The effective quark masses and coupling parameters are extracted from the conventional hadron masses \cite{ParticleDataGroup:2022pth}. One can find details about the extraction procedure in Refs. \cite{Wu:2018xdi,Wu:2016gas,Li:2023wug}. The quark masses that we obtained are $m_c=1724.1$ MeV, $m_b=5054.4$ MeV, $m_n=361.8$ MeV, and $m_s=542.4$ MeV. The coupling parameters are listed in Table \ref{coupling}. Note that the quark masses will be adopted only when the upper limits for the tetraquark masses with Eq. \eqref{Mass} are estimated.

Our results from Eq. \eqref{mass4140} rely on the effective quark mass gaps $\Delta_{cs}$, $\Delta_{sn}$, $\Delta_{bs}$, $\Delta_{bc}$, and $\Delta_{cn}$, which can also be extracted from the conventional hadron masses. The values $\Delta_{bc}=3340.2$ MeV, $\Delta_{cn}=1280.7$ MeV, and $\Delta_{sn}=90.6$ MeV were fixed in Refs. \cite{Cheng:2020nho,Wu:2018xdi}. Table \ref{QMD} shows the extracted $\Delta_{cs}$ and $\Delta_{bs}$ by using various hadrons. We adopt $\Delta_{cs}= 1180.6$ MeV and $\Delta_{bs}= 4520.2$ MeV in our calculation. The $X(4140)$ mass is taken to be 4146.5 MeV \cite{ParticleDataGroup:2022pth}. The corresponding CMI eigenvalue is $-85.5$ MeV, and then we obtain $\tilde{m}=4232.0$ MeV. To estimate the lower limits for the tetraquark masses with Eq. \eqref{massref} and calculate the decay widths, we also need the following meson masses: $M(D)=1867.2$ MeV, $M(D^*)=2008.6$ MeV, $M(D_s)=1968.3$ MeV, $M(D_s^*)=2112.2$ MeV, $M(\eta_c)=2983.9$ MeV, $M(J/\psi)=3096.9$ MeV, $M(B)=5279.5$ MeV, $M(B^*)=5324.7$ MeV, $M(B_s)=5366.9$ MeV, $M(B_s^*)=5415.4$ MeV, $M(\eta_b)=9399.0$ MeV, $M(\Upsilon)=9460.3$ MeV, $M(B_c)=6274.9$ MeV, and $M(B_c^*)=6344.9$ MeV. Note that the mass of the undiscovered $B_c^*$ is calculated within the CMI model.

\begin{table}[htbp]
\caption{Quark mass gaps $\Delta_{cs}$ and $\Delta_{bs}$ (units: MeV) determined from various conventional hadron masses.}\label{QMD}
\centering
\begin{tabular}{ccc|ccc}\hline\hline
Hadron&Hadron&$\Delta_{cs}$&Hadron&Hadron&$\Delta_{bs}$\\\hline
$J/\psi$&$\phi$&1049.4&$\Upsilon$&$\phi$&4237.5\\
$J/\psi(\eta_c)$&$D_{s}^*(D_s)$&992.2 (993.2)&$\Upsilon(\eta_b)$&$B_{s}^*(B_s)$&4041.7 (4041.8)\\
$D^*(D)$&$K^*(K)$&1180.6 (1179.4)&$B^*(B)$&$K^*(K)$&4520.2 (4518.8)\\
$D_s$&$\phi$&1106.6&$B_s$&$\phi$&4433.8\\
$B_c$&$B_s$&924.1&$B_c$&$D_s$&4252.2\\
$\Lambda_c$&$\Lambda$&1170.8&$\Lambda_b$&$\Lambda$&4503.8\\
$\Sigma_c^*(\Sigma_c)$&$\Sigma^*(\Sigma)$&1176.2 (1178.4)&$\Sigma_b^*(\Sigma_b)$&$\Sigma^*(\Sigma)$&4506.1 (4509.5)\\
$\Xi_c^*(\Xi_c^\prime)$&$\Xi^*(\Xi)$&1137.3 (1159.1)&$\Xi_b^*(\Xi_b^\prime)$&$\Xi^*(\Xi)$&4463.2 (4483.7)\\
$\Omega_c^*$&$\Omega$&1100.3&$\Omega_b$&$\Omega$&4415.5\\
$\Xi_{cc}$&$\Xi$&1112.2\\
\hline\hline
\end{tabular}
\end{table}

Even though we employ an oversimplified scheme to study the rearrangement decays, we still encounter the problem of determining the value of constant ${\cal C}$ because it may be different from system to system and its determined value depends on the considered decay channels. For the $cs\bar{c}\bar{s}$ system, ${\cal C}$ is around 7.3 GeV \cite{Li:2023wxm} when the assumption that the total decay width of $X(4140)$ is equal to the sum of the partial widths of the rearrangement decay channels is used. For the $cc\bar{c}\bar{c}$ case, ${\cal C}$ is around 15 GeV if $X(6600)$ is treated as the ground scalar tetraquark and $M=6552$ MeV and $\Gamma_{total}=124$ MeV \cite{Zhang:2022toq} values are used, along with a similar decay assumption. At present, since no triply heavy tetraquark candidate has been observed, we just present the width results with ${\cal C}=14954.7$ MeV.
	
\begin{figure}[htbp]\centering
\begin{tabular}{ccc}	\includegraphics[width=150pt]{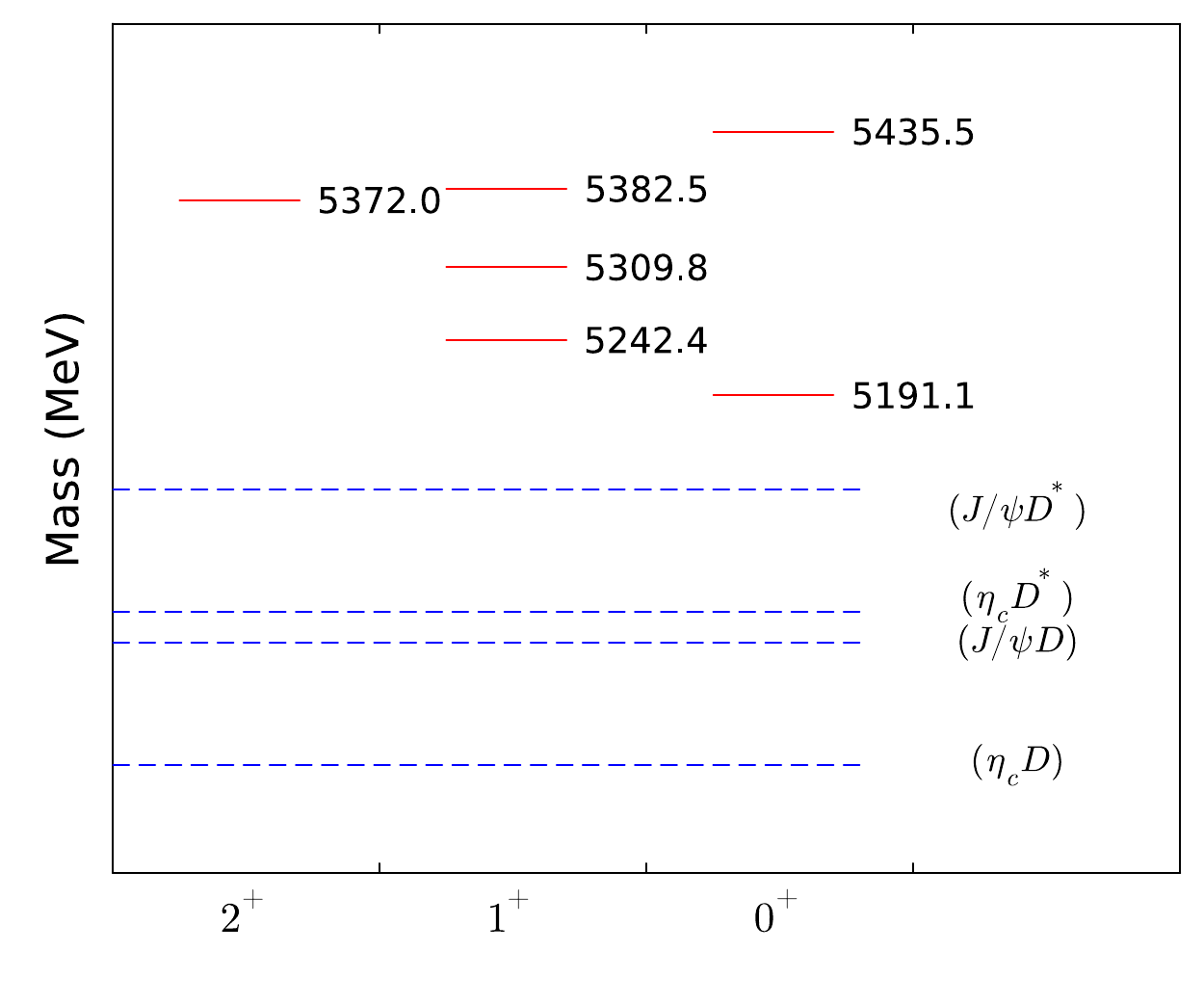}&\includegraphics[width=150pt]{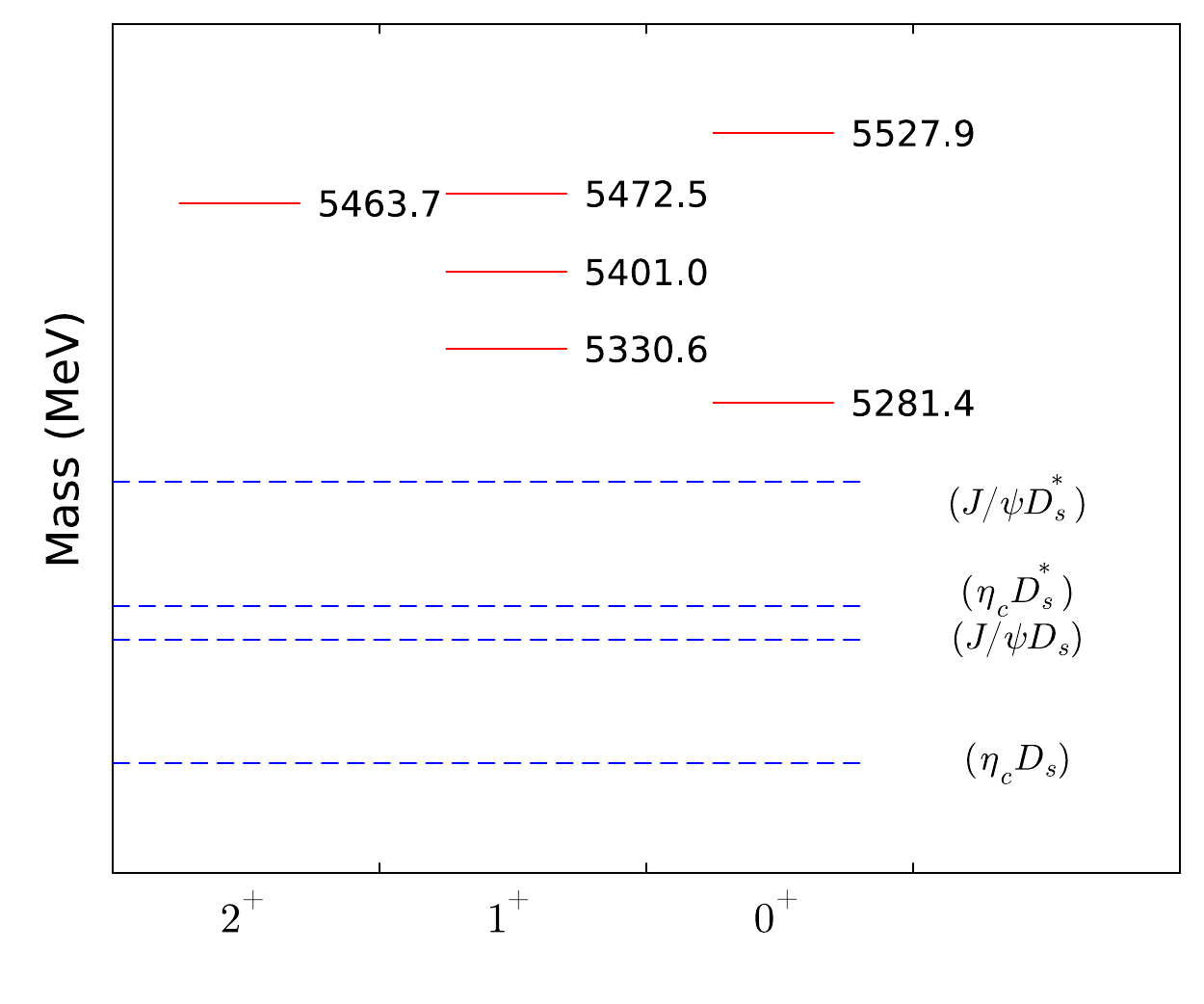}&\includegraphics[width=150pt]{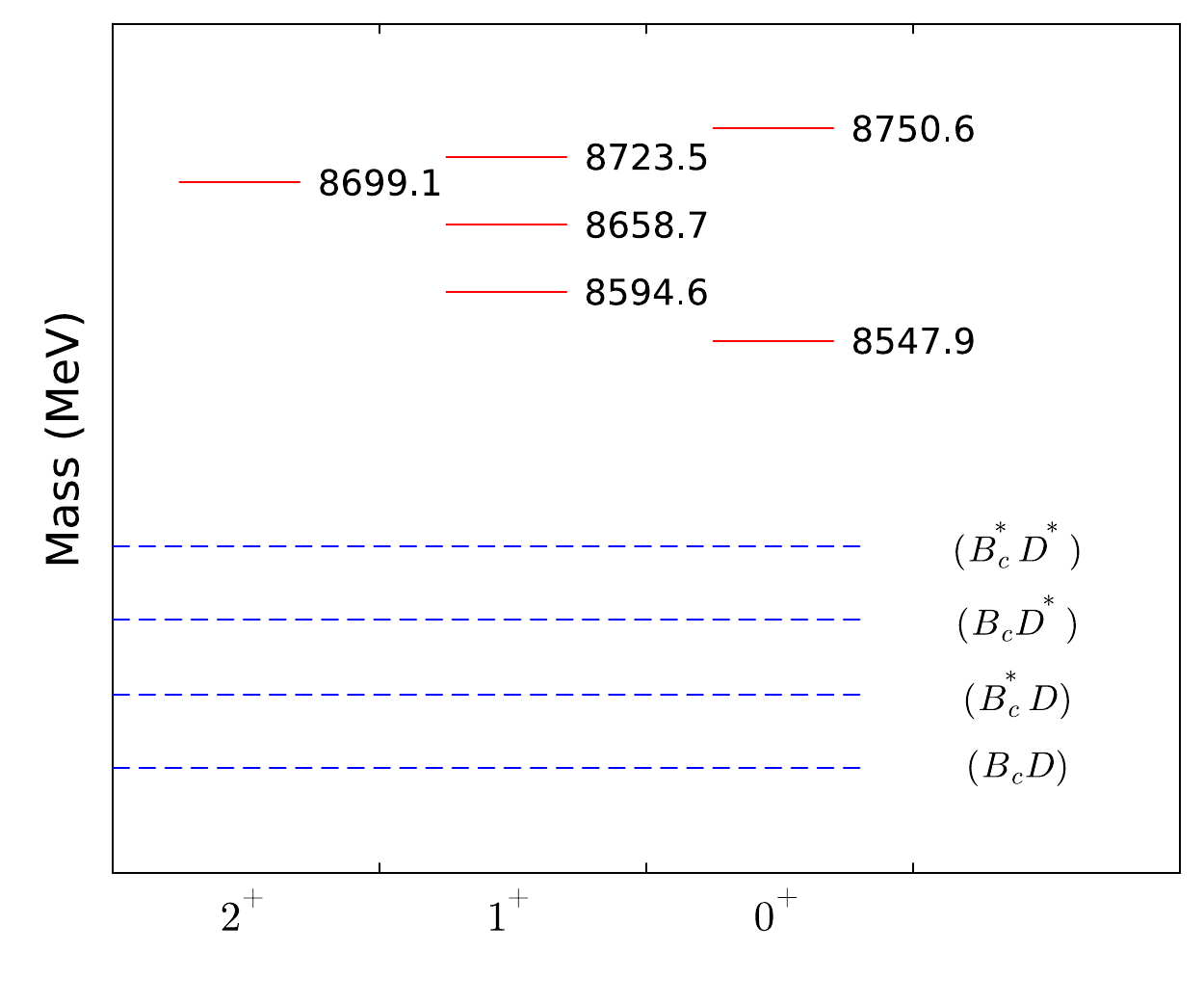}\\
(a) $cc\bar{c}\bar{n}$ states &  (b) $cc\bar{c}\bar{s}$ states & (c) $cc\bar{b}\bar{n}$ states\\
\includegraphics[width=150pt]{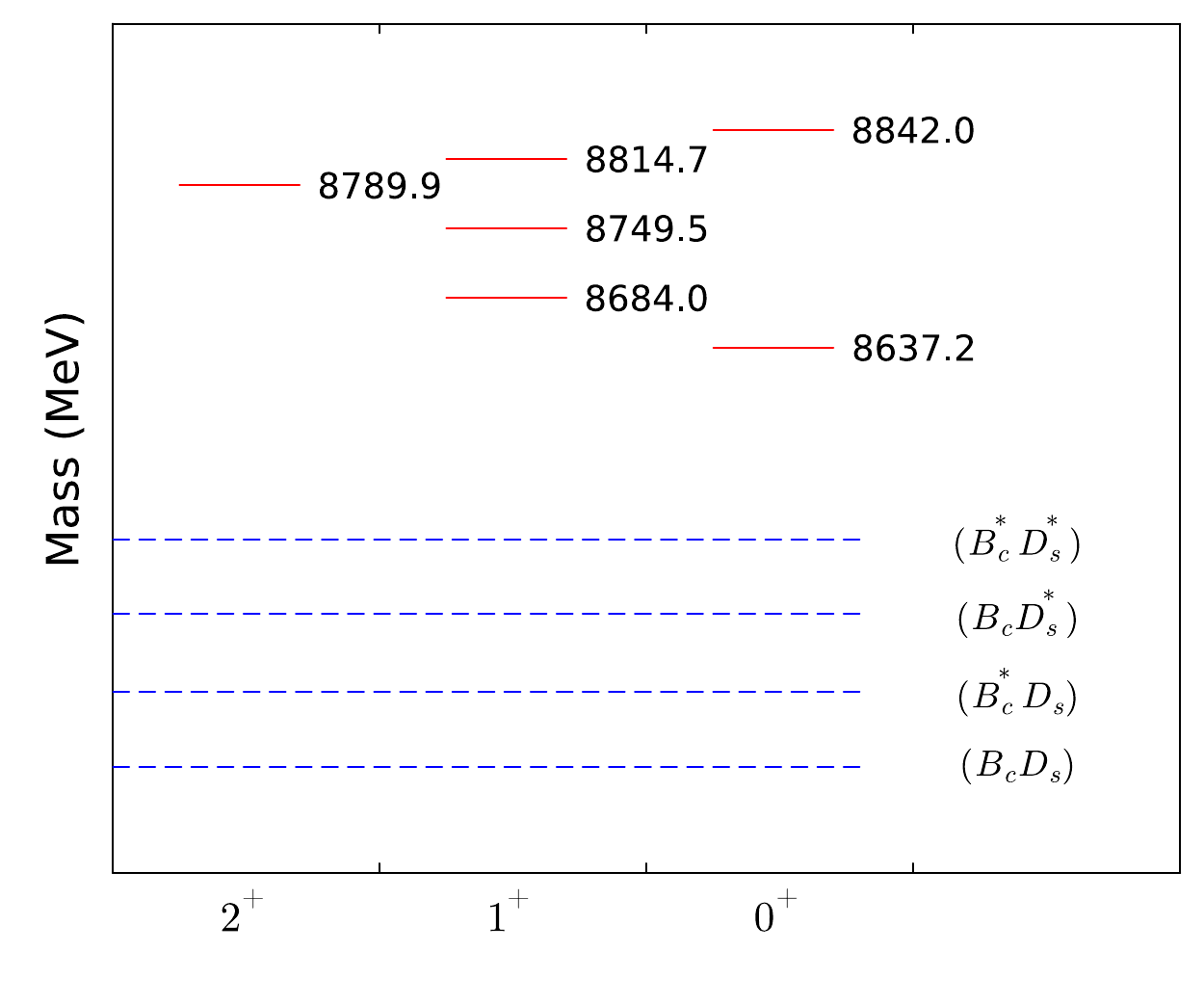}&\includegraphics[width=150pt]{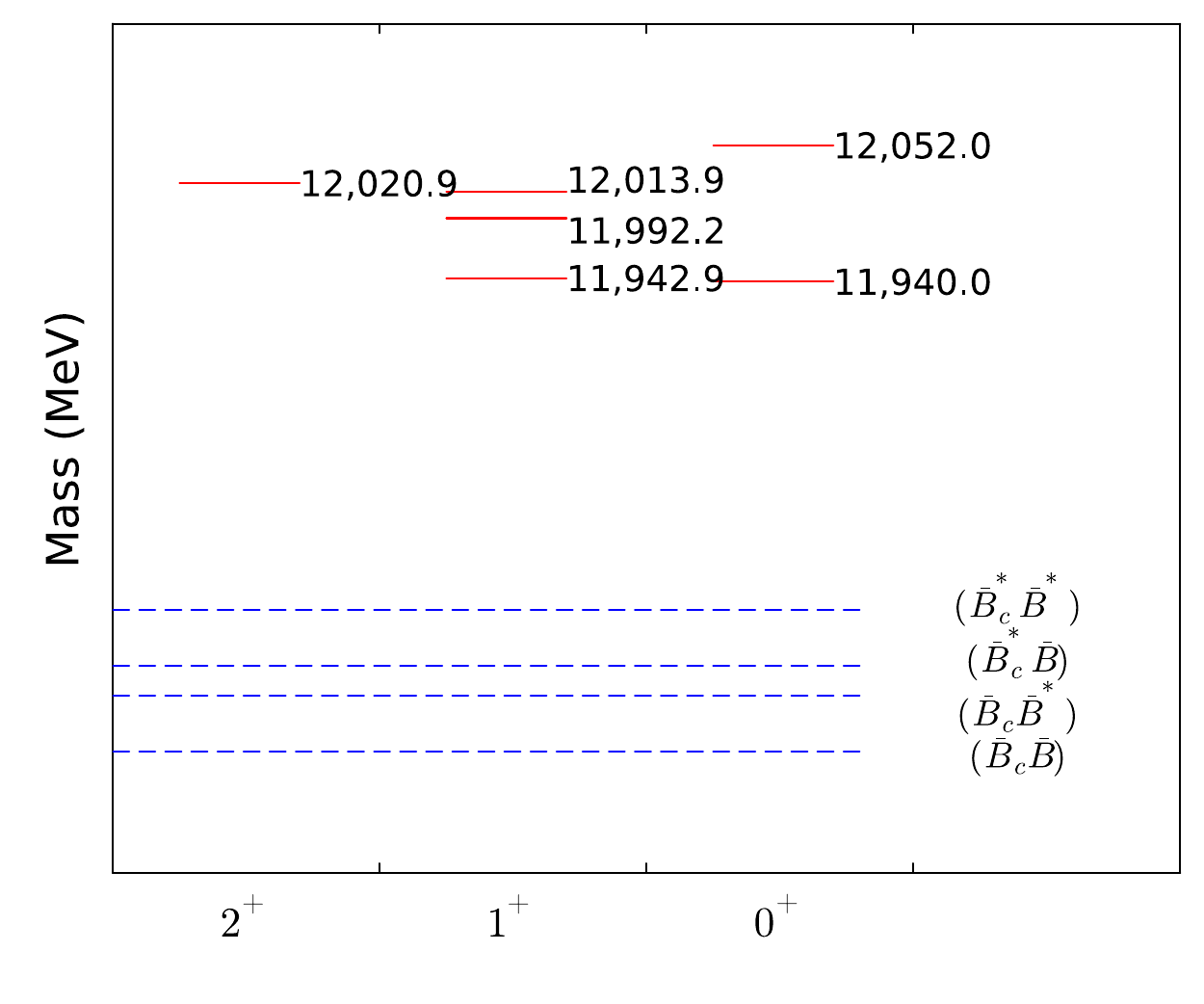}&\includegraphics[width=150pt]{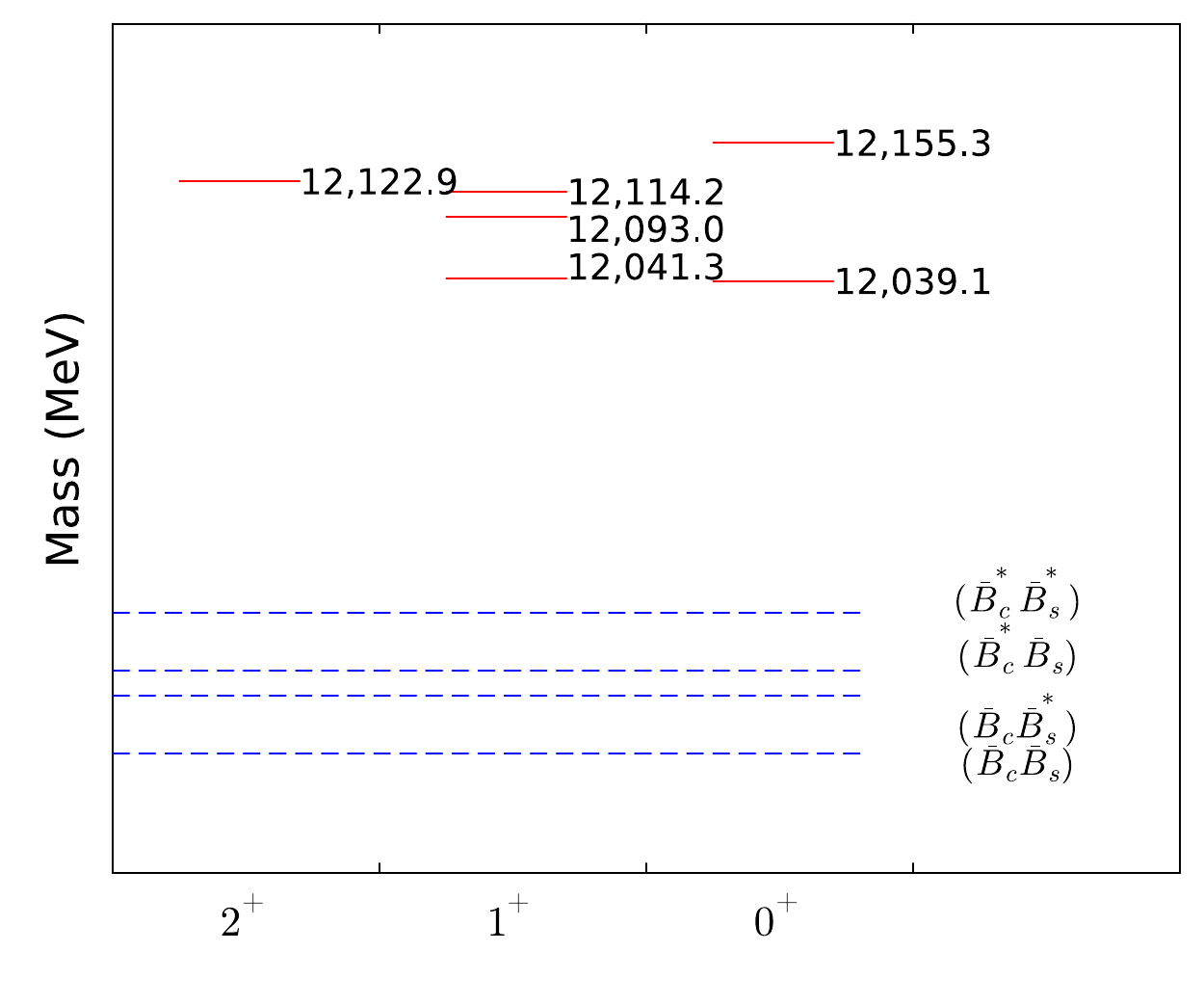}\\
(d) $cc\bar{b}\bar{s}$ states &  (e) $bb\bar{c}\bar{n}$ states & (f) $bb\bar{c}\bar{s}$ states\\
\includegraphics[width=150pt]{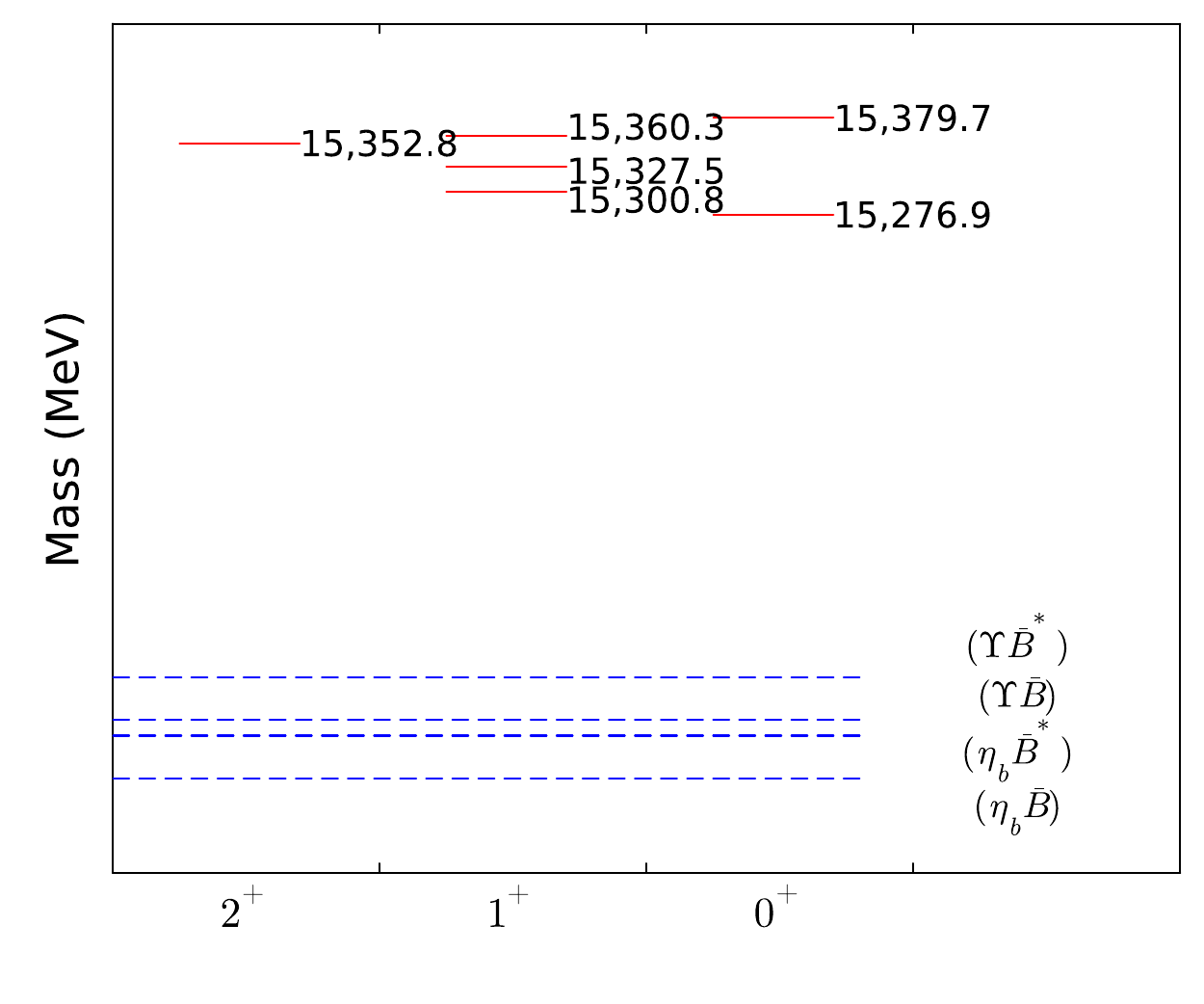}&\includegraphics[width=150pt]{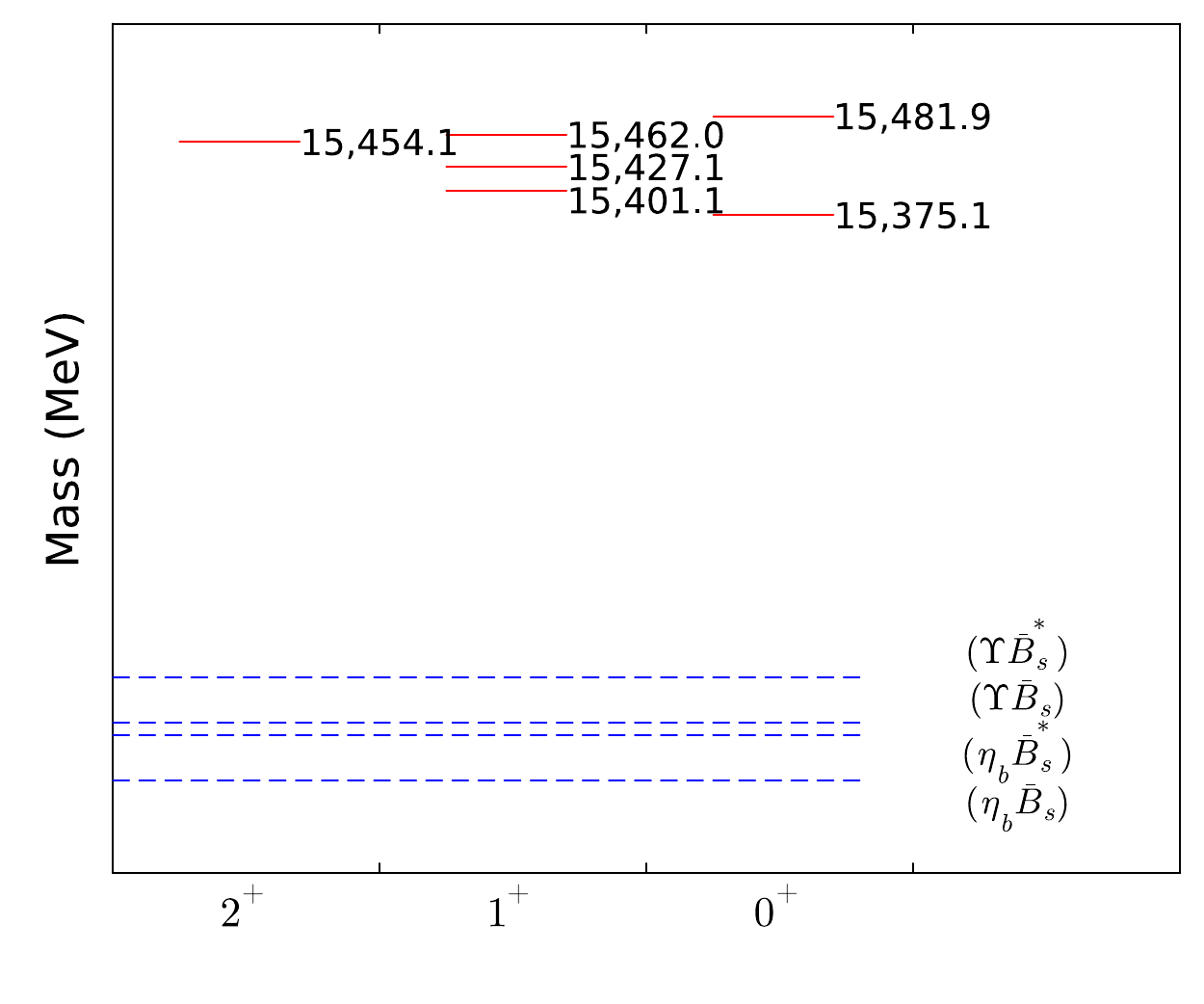}&\includegraphics[width=150pt]{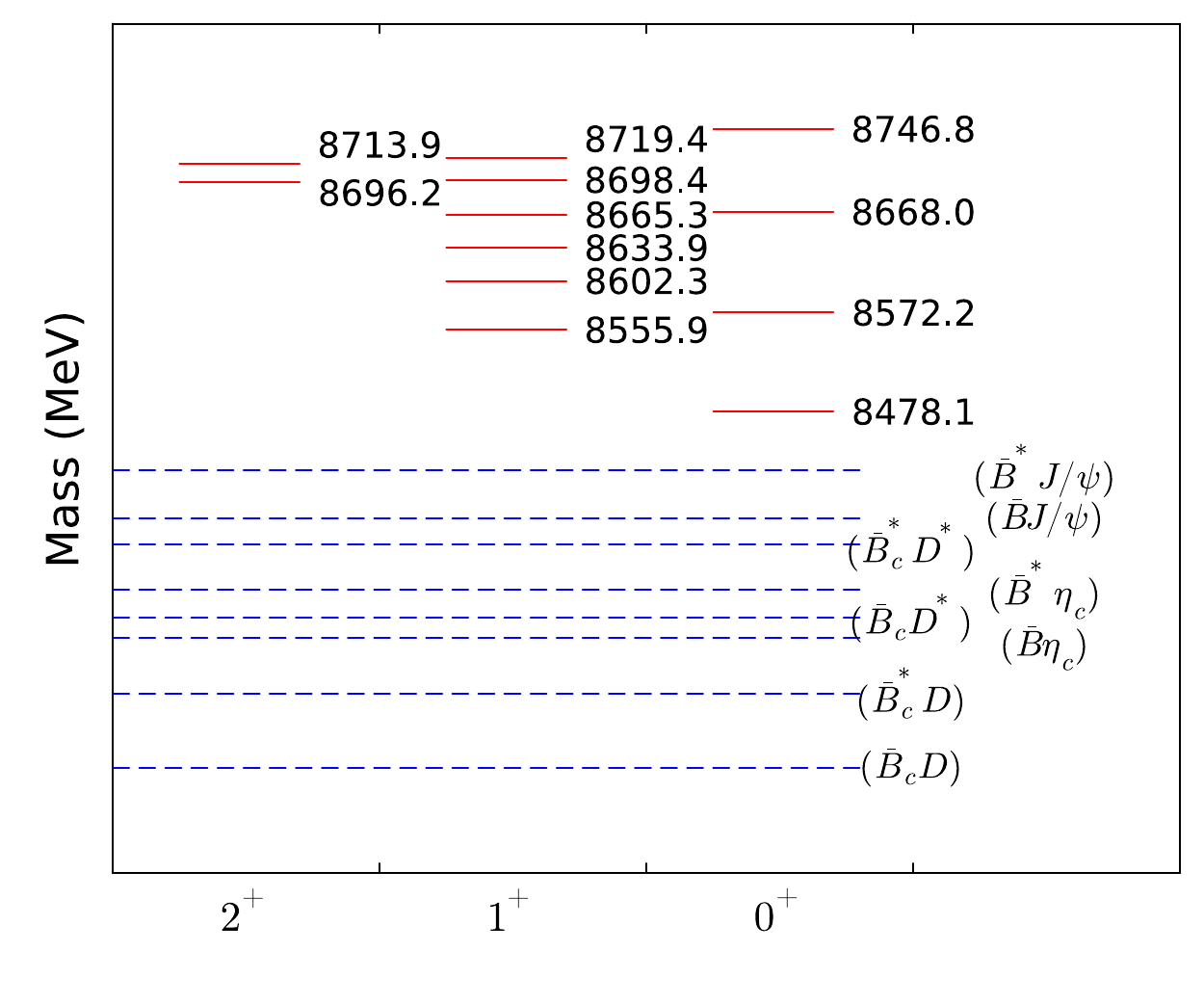}\\
(g) $bb\bar{b}\bar{n}$ states &  (h) $bb\bar{b}\bar{s}$ states & (i) $bc\bar{c}\bar{n}$ states\\
\includegraphics[width=150pt]{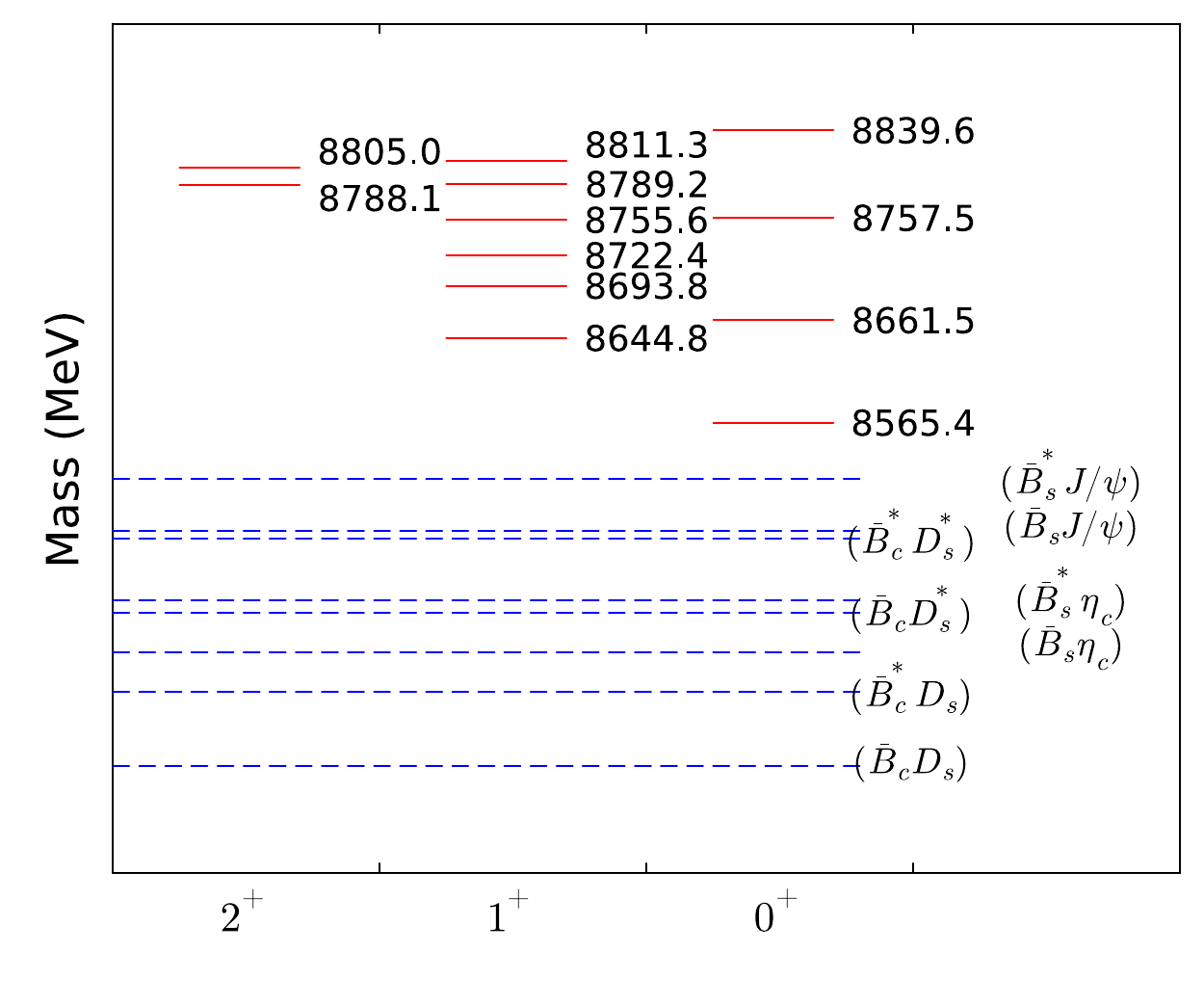}&\includegraphics[width=150pt]{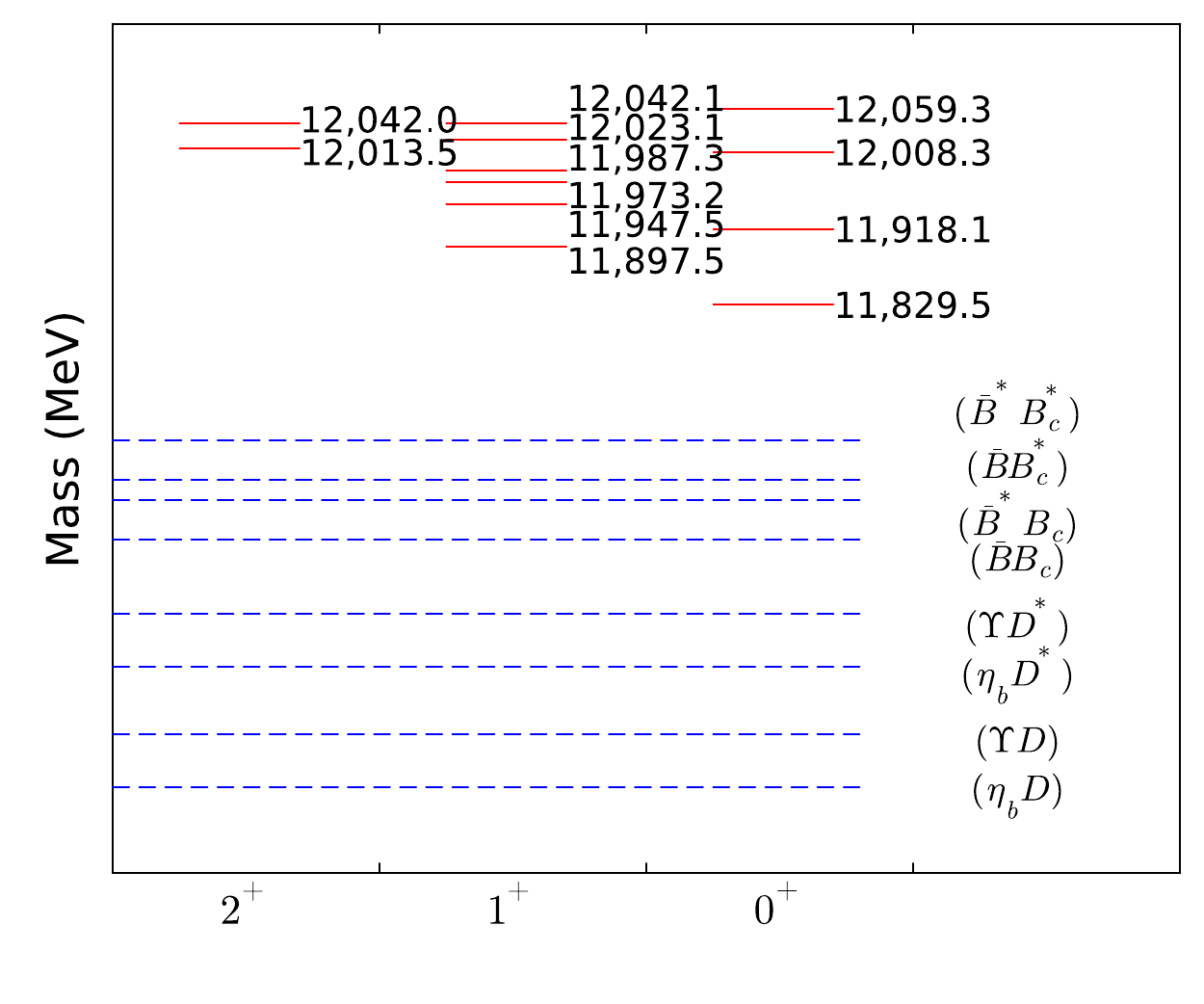}&\includegraphics[width=150pt]{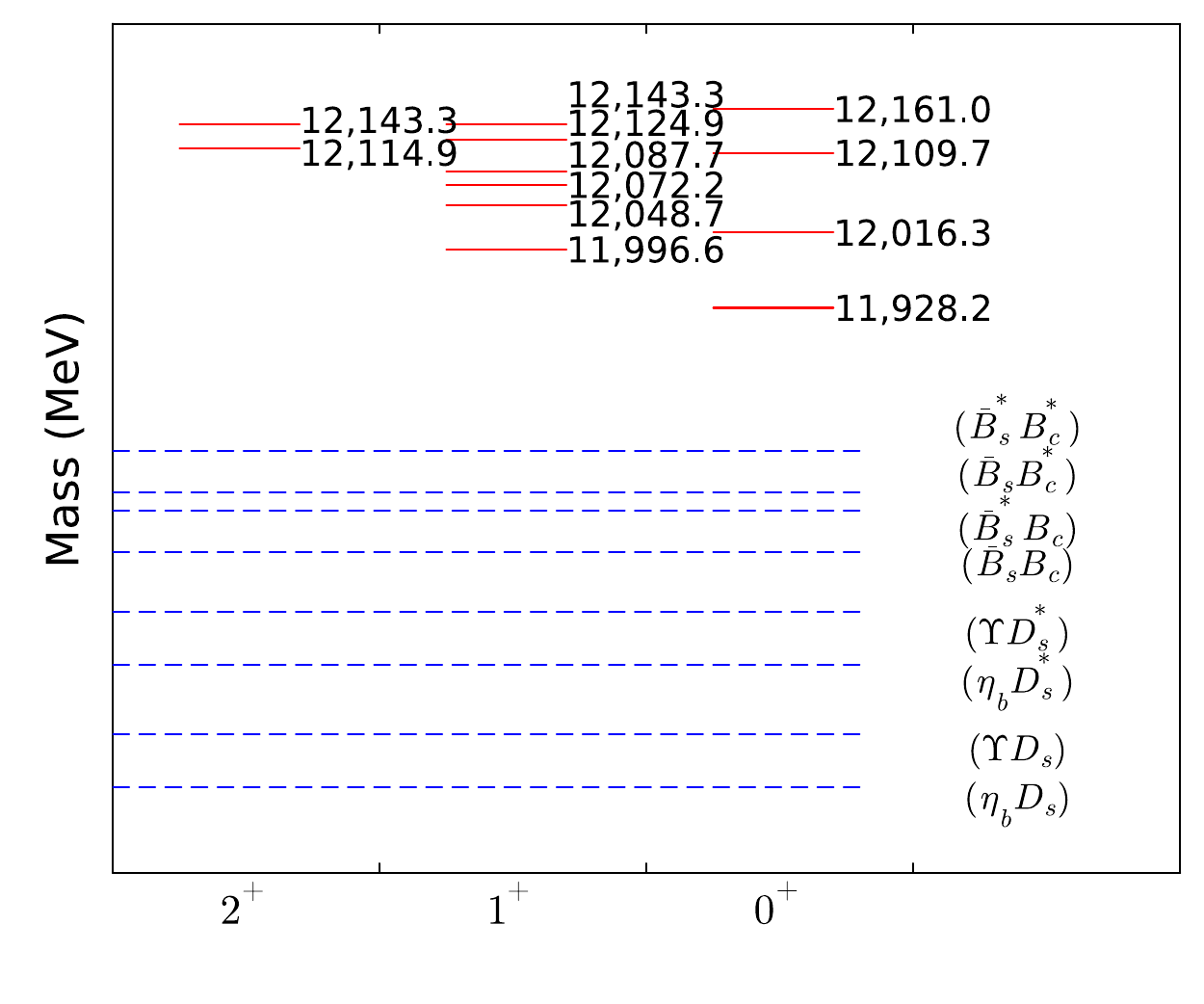}\\
(j) $bc\bar{c}\bar{s}$ states &  (k) $bc\bar{b}\bar{n}$ states & (l) $bc\bar{b}\bar{s}$ states
\end{tabular}
\caption{Relative positions of the triply heavy tetraquark states and their rearrangement decay channels.}\label{triply figures}
\end{figure}

With the above determined parameters, the numerical results for the mass spectra and rearrangement decay widths of the triply heavy tetraquarks $cc\bar{Q}\bar{q}$, $bb\bar{Q}\bar{q}$, and $bc\bar{Q}\bar{q}$ could be calculated. We display the relative positions for all the considered states in Fig. \ref{triply figures}. Details about their masses and widths are listed in tables \ref{mass of ccQq}--\ref{decay of bcQq}.

\subsection{The $cc\bar{c}\bar{n}$, $cc\bar{c}\bar{s}$, $cc\bar{b}\bar{n}$, and $cc\bar{b}\bar{s}$ states}

The CMI matrices and their eigenvalues and corresponding eigenvectors, as well as the masses calculated using Eqs. \eqref{Mass}-\eqref{mass4140}, for these four tetraquark systems are given in table \ref{mass of ccQq}. The masses calculated using $X(4140)$ as the reference state are our predicted values. The relative positions of these $cc\bar{Q}\bar{q}$ states and related decay channels are illustrated Fig. \ref{triply figures}(a)-(d). For the lower limits calculation, several meson-meson thresholds could be adopted. We use the thresholds of $J/\psi D$, $J/\psi D_s$, $B_c D$, and $B_c D_s$ for $cc\bar{c}\bar{n}$, $cc\bar{c}\bar{s}$, $cc\bar{b}\bar{n}$, and $cc\bar{b}\bar{s}$, respectively. Other choices do not affect much. The tetraquark mass upper and lower limits in table \ref{mass of ccQq} are consistent with the results in Ref. \cite{Chen:2016ont}, but the obtained masses used in the following discussions are between the upper and lower limits. Due to the values being larger than those in Ref. \cite{Chen:2016ont}, we find no stable tetraquarks in these systems, while stable $2^+$ $cc\bar{Q}\bar{q}$ states are found to be possible in Ref. \cite{Chen:2016ont}. The relevant rearrangement decay channels, corresponding partial widths, and total widths are listed in table \ref{decay of ccQq}.

The $cc\bar{c}\bar{n}$ system has six states, whose masses range from 5191.1 MeV to 5435.5 MeV. The highest and lowest tetraquarks are both $J^P=0^+$ states and they can decay into the same channels: $J/\psi D^*$ and $\eta_c D$. The rearrangement decays of the higher and lower $0^+$ state are dominated by the $J/\psi D^*$ and $\eta_c D$ channels, respectively. There are three $1^+$ tetraquarks located from 5242.4 to 5382.5 MeV, with $J/\psi D^*$, $J/\psi D$, and $\eta_c D^*$ as their rearrangement decay modes. The highest state mainly decays into $J/\psi D^*$. Both the intermediate and lowest states have two dominant channels, namely, $J/\psi D$ and $\eta_c D^*$, but the $\Gamma_{J/\psi D}:\Gamma_{\eta_c D^*}$ ratios are different. Their predicted values are about 0.5 and 1.8, respectively. The $2^+$ state is located at 5372.0 MeV, with only one decay mode $J/\psi D^*$, when only the $S$-wave channel is considered. When the $D$-wave decays are also counted, it can also decay into $\eta_cD^*$, $J/\psi D$, and $\eta_cD$.

The $cc\bar{b}\bar{n}$ tetraquarks span a range from 8547.9 to 8750.6 MeV, with the $J^P$ of the highest and lowest states being $0^+$. The two $0^+$ tetraquarks have $B_c^* D^*$ and $B_c D$ decay channels. The higher state has a dominant $B_c^* D^*$ decay mode, while the lower state mainly decays into $B_c D$. Three $1^+$ tetraquarks are located in the range of 8594.6$\sim$8723.5 MeV. The dominant decay channels for the highest, intermediate, and lowest states are $B_c^* D^*$, $B_c D^*$, and $B_c^* D$, respectively. The $2^+$ state is located around 8700 MeV and should mainly decay into $B_c^* D^*$ via an $S$-wave interaction, but the other three suppressed $D$-wave channels $B_cD^*$, $B_c^*D$, and $B_cD$ are also allowed.
	
When comparing the results for the $cc\bar{c}\bar{s}$ states with those for $cc\bar{c}\bar{n}$, we find that they have almost the same eigenvalues, eigenvectors, and decay information. The $cc\bar{c}\bar{s}$ tetraquarks are about 90 MeV higher than the corresponding $cc\bar{c}\bar{n}$ states, while the decay widths of $cc\bar{c}\bar{s}$ are slightly smaller than those of the corresponding $cc\bar{c}\bar{n}$. The above features are also observed in the $cc\bar{b}\bar{n}$ and $cc\bar{b}\bar{s}$ cases. For the $cc\bar{c}\bar{s}$ ($cc\bar{b}\bar{s}$) tetraquarks, the dominant decay channels are nearly the same as those of $cc\bar{c}\bar{n}$ ($cc\bar{b}\bar{n}$), except for replacing $D^{(*)}$ in the final states with $D_s^{(*)}$.

\begin{table}[htbp]
\caption{Numerical results for the $cc\bar{Q}\bar{q}$ systems in MeV units. The lower limits for the tetraquark masses in the seventh column are calculated using the reference meson-meson states $J/\psi D$, $J/\psi D_s$, $B_cD$, and $B_cD_s$ for the  $cc\bar{c}\bar{n}$, $cc\bar{c}\bar{s}$, $cc\bar{b}\bar{n}$, and $cc\bar{b}\bar{s}$ systems, respectively. The tetraquark masses in the sixth column are calculated by using $X(4140)$ as the reference state. The corresponding $\langle H_{CMI}\rangle$ base vectors are given in Sec. \ref{basevector}.}\label{mass of ccQq}\scriptsize
\begin{tabular}{cccccccc}
\hline
System&$J^{P}$ & $\langle H_{CMI}\rangle$ &Eigenvalue &Eigenvector &Mass& Lower limit &Upper limit\\\hline
$cc\bar{c}\bar{n}$&$2^{+}$ &$\left(\begin{array}{c}50.9\end{array}\right)$&$\left(\begin{array}{c}50.9\end{array}\right)$&$\left[\begin{array}{c}\{1.00\}\end{array}\right]$&$\left(\begin{array}{c}5372.0\end{array}\right)$&$\left(\begin{array}{c}5092.4\end{array}\right)$&$\left(\begin{array}{c}5585.0\end{array}\right)$\\
&$1^{+}$ &$\left(\begin{array}{ccc}-12.5&-4.9&-10.4\\-4.9&-23.5&-67.3\\-10.4&-67.3&7.5\end{array}\right)$&$\left(\begin{array}{c}61.4\\-11.3\\-78.7\end{array}\right)$&$\left[\begin{array}{ccc}\{-0.07,-0.62,0.78\}\\\{0.99,-0.16,-0.04\}\\\{-0.15,-0.77,-0.62\}\end{array}\right]$&$\left(\begin{array}{c}5382.5\\5309.8\\5242.4\end{array}\right)$&$\left(\begin{array}{c}5102.9\\5030.2\\4962.8\end{array}\right)$&$\left(\begin{array}{c}5595.5\\5522.8\\5455.4\end{array}\right)$\\
&$0^{+}$ &$\left(\begin{array}{cc}-44.3&116.6\\116.6&28.8\end{array}\right)$&$\left(\begin{array}{c}114.5\\-129.9\end{array}\right)$&$\left[\begin{array}{cc}\{-0.59,-0.81\}\\\{-0.81,0.59\}\end{array}\right]$&$\left(\begin{array}{c}5435.5\\5191.1\end{array}\right)$&$\left(\begin{array}{c}5155.9\\4911.5\end{array}\right)$&$\left(\begin{array}{c}5648.6\\5404.2\end{array}\right)$\\
\hline
$cc\bar{c}\bar{s}$&$2^{+}$ &$\left(\begin{array}{c}52.0\end{array}\right)$&$\left(\begin{array}{c}52.0\end{array}\right)$&$\left[\begin{array}{c}\{1.00\}\end{array}\right]$&$\left(\begin{array}{c}5463.7\end{array}\right)$&$\left(\begin{array}{c}5196.1\end{array}\right)$&$\left(\begin{array}{c}5766.7\end{array}\right)$\\
  &$1^{+}$ &$\left(\begin{array}{ccc}-12.0&-5.3&-11.2\\-5.3&-25.9&-67.9\\-11.2&-67.9&7.1\end{array}\right)$&$\left(\begin{array}{c}60.9\\-10.6\\-81.0\end{array}\right)$&$\left[\begin{array}{ccc}\{-0.08,-0.61,0.79\}\\\{0.98,-0.17,-0.04\}\\\{-0.16,-0.77,-0.62\}\end{array}\right]$&$\left(\begin{array}{c}5472.5\\5401.0\\5330.6\end{array}\right)$&$\left(\begin{array}{c}5205.0\\5133.5\\5063.1\end{array}\right)$&$\left(\begin{array}{c}5775.6\\5704.1\\5633.7\end{array}\right)$\\
 &$0^{+}$ &$\left(\begin{array}{cc}-44.0&117.6\\117.6&30.0\end{array}\right)$&$\left(\begin{array}{c}116.3\\-130.3\end{array}\right)$&$\left[\begin{array}{cc}\{-0.59,-0.81\}\\\{-0.81,0.59\}\end{array}\right]$&$\left(\begin{array}{c}5527.9\\5281.4\end{array}\right)$&$\left(\begin{array}{c}5260.4\\5013.9\end{array}\right)$&$\left(\begin{array}{c}5831.0\\5584.4\end{array}\right)$\\
\hline
$cc\bar{b}\bar{n}$&$2^{+}$ &$\left(\begin{array}{c}38.4\end{array}\right)$&$\left(\begin{array}{c}38.4\end{array}\right)$&$\left[\begin{array}{c}\{1.00\}\end{array}\right]$&$\left(\begin{array}{c}8699.1\end{array}\right)$&$\left(\begin{array}{c}8338.9\end{array}\right)$&$\left(\begin{array}{c}8902.8\end{array}\right)$\\
&$1^{+}$ &$\left(\begin{array}{ccc}-14.4&-12.4&-26.4\\-12.4&-1.9&-56.0\\-26.4&-56.0&11.1\end{array}\right)$&$\left(\begin{array}{c}62.8\\-2.0\\-66.0\end{array}\right)$&$\left[\begin{array}{ccc}\{-0.16,-0.63,0.76\}\\\{0.87,-0.45,-0.19\}\\\{-0.47,-0.63,-0.62\}\end{array}\right]$&$\left(\begin{array}{c}8723.5\\8658.7\\8594.6\end{array}\right)$&$\left(\begin{array}{c}8363.3\\8298.5\\8234.5\end{array}\right)$&$\left(\begin{array}{c}8927.2\\8862.4\\8798.4\end{array}\right)$\\
 & $0^{+}$ &$\left(\begin{array}{cc}-40.8&97.0\\97.0&18.0\end{array}\right)$&$\left(\begin{array}{c}90.0\\-112.8\end{array}\right)$&$\left[\begin{array}{cc}\{-0.60,-0.80\}\\\{-0.80,0.60\}\end{array}\right]$&$\left(\begin{array}{c}8750.6\\8547.9\end{array}\right)$&$\left(\begin{array}{c}8390.5\\8187.7\end{array}\right)$&$\left(\begin{array}{c}8954.4\\8751.6\end{array}\right)$\\
\hline
$cc\bar{b}\bar{s}$&$2^{+}$ &$\left(\begin{array}{c}38.7\end{array}\right)$&$\left(\begin{array}{c}38.7\end{array}\right)$&$\left[\begin{array}{c}\{1.00\}\end{array}\right]$&$\left(\begin{array}{c}8789.9\end{array}\right)$&$\left(\begin{array}{c}8441.9\end{array}\right)$&$\left(\begin{array}{c}9083.7\end{array}\right)$\\
 &$1^{+}$ &$\left(\begin{array}{ccc}-14.7&-12.8&-27.2\\-12.8&-1.9&-56.6\\-27.2&-56.6&11.1\end{array}\right)$&$\left(\begin{array}{c}63.5\\-1.7\\-67.2\end{array}\right)$&$\left[\begin{array}{ccc}\{-0.16,-0.63,0.76\}\\\{0.87,-0.46,-0.20\}\\\{-0.47,-0.63,-0.62\}\end{array}\right]$&$\left(\begin{array}{c}8814.7\\8749.5\\8684.0\end{array}\right)$&$\left(\begin{array}{c}8466.7\\8401.5\\8336.0\end{array}\right)$&$\left(\begin{array}{c}9108.5\\9043.3\\8977.8\end{array}\right)$\\
   &$0^{+}$ &$\left(\begin{array}{cc}-41.3&98.0\\98.0&18.0\end{array}\right)$&$\left(\begin{array}{c}90.7\\-114.0\end{array}\right)$&$\left[\begin{array}{cc}\{-0.60,-0.80\}\\\{-0.80,0.60\}\end{array}\right]$&$\left(\begin{array}{c}8842.0\\8637.2\end{array}\right)$&$\left(\begin{array}{c}8493.9\\8289.2\end{array}\right)$&$\left(\begin{array}{c}9135.7\\8931.0\end{array}\right)$\\
\hline
\end{tabular}
\end{table}
	
\begin{table}[htbp]
\centering
\caption{Rearrangement decays for the $cc\bar{c}\bar{n}$, $cc\bar{c}\bar{s}$, $cc\bar{b}\bar{n}$, and $cc\bar{b}\bar{s}$ states. The numbers in the parentheses are $(100|\mathcal{M}|^2/{\cal C}^2, \Gamma)$. The tetraquark mass, partial width $\Gamma$, and total width $\Gamma_{sum}$ values are given in MeV units.}\label{decay of ccQq}\scriptsize
\begin{tabular}{ccccccc}
\hline\hline
System&$J^P$ & Mass & \multicolumn{3}{c}{ Decay channels} & $\Gamma_{sum}$ \\ \hline
$cc\bar{c}\bar{n}$ && & $J/\psi D^*$ & & & \\
 &$2^+$ & $\left[\begin{array}{c} 5372.0\end{array} \right]$ & $\left[\begin{array}{c}(33.3, 168.2)  \end{array}\right]$ & & & $\left[\begin{array}{c}168.2  \end{array}\right]$ \\
   &&&$J/\psi D^*$&$J/\psi D$&$\eta_c D^*$& \\
&$1^+$ & $\left[\begin{array}{c}5382.5\\5309.8\\5242.4 \end{array}\right]$&  $\left[\begin{array}{c}(49.6, 254.5)\\ (0.2, 0.8) \\ (0.2, 0.7) \end{array}\right]$ &  $\left[\begin{array}{c}(1.3, 8.0) \\ (11.4, 66.0) \\ (29.0, 153.6) \end{array}\right]$ &  $\left[\begin{array}{c}(2.9, 17.6) \\ (21.8, 122.3) \\ (17.0, 86.4) \end{array}\right]$ &  $\left[\begin{array}{c}280.0\\189.1\\240.8 \end{array}\right]$\\
   &&&$J/\psi D^*$&$\eta_c D$&&\\
&$0^+$& $\left[\begin{array}{c}5435.5\\5191.1 \end{array}\right]$& $\left[\begin{array}{c}(54.9, 302.0) \\ (3.5, 10.5) \end{array}\right]$& $\left[\begin{array}{c}(0.1, 0.8)\\ (41.6, 247.6) \end{array}\right]$& & $\left[\begin{array}{c}302.8\\ 258.1 \end{array}\right]$\\
\hline
$cc\bar{c}\bar{s}$&&&$J/\psi D_s^*$&&&\\
  &$2^+$& $\left[\begin{array}{c}5463.7 \end{array}\right]$& $\left[\begin{array}{c}(33.3, 161.1) \end{array}\right]$&&& $\left[\begin{array}{c}161.1 \end{array}\right]$\\
   &&&$J/\psi D_s^*$&$J/\psi D_s$&$\eta_c D_s^*$& \\
&$1^+$& $\left[\begin{array}{c}5472.5\\5401.0\\5330.6 \end{array}\right]$& $\left[\begin{array}{c}(49.6, 243.2) \\ (0.2, 1.0)\\ (0.2, 0.6) \end{array}\right]$& $\left[\begin{array}{c}(1.3, 7.8)\\ (11.3, 63.4)\\ (29.0, 147.6) \end{array}\right]$& $\left[\begin{array}{c}(3.1, 18.2)\\ (21.8, 117.5)\\ (16.7, 80.9) \end{array}\right]$& $\left[\begin{array}{c}269.2\\181.9\\229.1 \end{array}\right]$\\
   &&&$J/\psi D_s^*$&$\eta_c D_s$&&\\
&$0^+$& $\left[\begin{array}{c}5527.9\\5281.4 \end{array}\right]$& $\left[\begin{array}{c}(54.9, 291.0)\\ (3.4, 9.4) \end{array}\right]$& $\left[\begin{array}{c}(0.1, 0.8)\\ (41.6, 238.8) \end{array}\right]$&& $\left[\begin{array}{c}291.8\\248.2 \end{array}\right]$\\
\hline
$cc\bar{b}\bar{n}$&&&$B_c^* D^*$&&&\\
&$2^+$& $\left[\begin{array}{c}8699.1 \end{array}\right]$& $\left[\begin{array}{c}(33.3, 82.5) \end{array}\right]$&&& $\left[\begin{array}{c}82.5 \end{array}\right]$\\
   &&&$B_c^* D^*$&$B_c^* D$&$B_c D^*$&\\
&$1^+$& $\left[\begin{array}{c}8723.5\\8658.7\\8594.6 \end{array}\right]$& $\left[\begin{array}{c}(48.5, 123.6)\\ (0.5, 1.3)\\ (1.0, 2.1) \end{array}\right]$& $\left[\begin{array}{c}(0.4, 1.2)\\ (2.1, 5.8)\\ (39.2, 102.1) \end{array}\right]$& $\left[\begin{array}{c}(3.8, 10.5)\\ (32.0, 83.2)\\ (5.9, 14.2) \end{array}\right]$& $\left[\begin{array}{c}135.4\\90.3\\118.4 \end{array}\right]$\\
   &&&$B_c^* D^*$&$B_c D$&&\\
&$0^+$& $\left[\begin{array}{c}8750.6\\8547.9 \end{array}\right]$& $\left[\begin{array}{c}(54.7, 144.0)\\ (3.6, 6.8) \end{array}\right]$& $\left[\begin{array}{c}(0.1, 0.3)\\ (41.6, 112.9) \end{array}\right]$&& $\left[\begin{array}{c}144.3\\119.7 \end{array}\right]$\\
\hline
$cc\bar{b}\bar{s}$&&&$B_c^* D_s^*$&&&\\
  &$2^+$& $\left[\begin{array}{c}8789.9 \end{array}\right]$& $\left[\begin{array}{c}(33.3, 80.6) \end{array}\right]$&&& $\left[\begin{array}{c}80.6 \end{array}\right]$\\
    &&&$B_c^* D_s^*$&$B_c^* D_s$&$B_c D_s^*$&\\
&$1^+$& $\left[\begin{array}{c}8814.7\\8749.5\\8684.0 \end{array}\right]$& $\left[\begin{array}{c}(48.4, 120.9)\\ (0.6, 1.3)\\ (1.0, 2.1) \end{array}\right]$& $\left[\begin{array}{c}(0.4, 1.2)\\ (2.0, 5.5)\\ (39.3, 100.5) \end{array}\right]$& $\left[\begin{array}{c}(3.8, 10.5)\\ (32.0, 81.7)\\ (5.8, 13.5) \end{array}\right]$& $\left[\begin{array}{c}132.6\\88.4\\116.0 \end{array}\right]$\\
    &&&$B_c^* D^*$&$B_c D$&&\\
&$0^+$& $\left[\begin{array}{c}8842.0\\8637.2 \end{array}\right]$& $\left[\begin{array}{c}(54.7, 141.1)\\ (3.6, 6.6) \end{array}\right]$& $\left[\begin{array}{c}(0.1, 0.3)\\ (41.6, 110.9) \end{array}\right]$&& $\left[\begin{array}{c}141.4\\117.4 \end{array}\right]$\\
\hline\hline
\end{tabular}
\end{table}

\subsection{The $bb\bar{c}\bar{n}$, $bb\bar{c}\bar{s}$, $bb\bar{b}\bar{n}$, and $bb\bar{b}\bar{s}$ states}

We list the CMI- and tetraquark-mass-related results for the $bb\bar{Q}\bar{q}$ states in table \ref{mass of bbQq}. The relative positions with masses using $X(4140)$ as the reference state are illustrated in Fig. \ref{triply figures}(e)-(h). For the lower limit calculations with Eq. \eqref{massref}, the meson-meson thresholds that we choose are those of $B_c^-\bar{B}$, $B_c^-\bar{B}^0_s$, $\Upsilon \bar{B}$, and $\Upsilon \bar{B}^0_s$ for $bb\bar{c}\bar{n}$, $bb\bar{c}\bar{s}$, $bb\bar{b}\bar{n}$, and $bb\bar{b}\bar{s}$, respectively. We obtain values of the lower limits (and upper limits with Eq. \eqref{Mass}) similar to those in \cite{Chen:2016ont}. The masses predicted in this study are between the lower and upper limits. Due to the higher spectra (see Fig. \ref{triply figures}), no stable tetraquark states exist in the $bb\bar{Q}\bar{q}$ systems. We present the relevant rearrangement decay channels, corresponding partial widths, and total widths in table \ref{decay of bbQq}.

The $bb\bar{c}\bar{n}$ tetraquarks are distributed in the $11940.0\sim12052.0$ MeV mass range, with the $J^P$ of both the highest and lowest states being $0^+$. The two $0^+$ states can both decay into $\bar{B}_c^*\bar{B}^*$ and $\bar{B}_c\bar{B}$ \footnote{Here we simply use $\bar{B}_c^{(*)}$ and $\bar{B}_s^{(*)}$ to denote $B_c^{(*)-}$ and $\bar{B}_s^{(*)0}$, respectively.}. The former (latter) channel is dominant for the higher (lower) tetraquark. There are three $1^+$ states that range from 11,942.9 to 12,013.9 MeV, with three decay channels: $\bar{B}_c^*\bar{B}^*$, $\bar{B}_c^*\bar{B}$, and $\bar{B}_c\bar{B}^*$. The respective partial width ratios of the dominant decay channels for the highest, intermediate, and lowest $1^+$ states are
\begin{equation}
\begin{aligned}
\Gamma_{\bar{B}_c^*\bar{B}^*}:\Gamma_{\bar{B}_c^*\bar{B}}&=4.5,\\
\Gamma_{\bar{B}_c^*\bar{B}^*}:\Gamma_{\bar{B}_c^*\bar{B}}:\Gamma_{\bar{B}_c\bar{B}^*}&=1.0:4.8:3.6,\\
\Gamma_{\bar{B}_c^*\bar{B}}:\Gamma_{\bar{B}_c\bar{B}^*}&=0.5.
\end{aligned}
\end{equation}
The $2^+$ state is located around 12.0 GeV, with only one $S$-wave decay channel: $\bar{B}_c^*\bar{B}^*$. The other three channels, namely, $\bar{B}_c^*\bar{B}$, $\bar{B}_c\bar{B}^*$, and $\bar{B}_c\bar{B}$, are also allowed if the $D$-wave decays are counted.

Like the discussion in the above $cc\bar{Q}\bar{q}$ case, we compared the $bb\bar{c}\bar{s}$ results with the $bb\bar{c}\bar{n}$ results and found similar features. Our estimation indicates that the $bb\bar{c}\bar{s}$ tetraquarks are about 100 MeV heavier than the $bb\bar{c}\bar{n}$ states, but they have comparable widths. By replacing $\bar{B}^{(*)}$ with $\bar{B}_s^{(*)}$ in the final states, we obtain the decay channels of the $bb\bar{c}\bar{s}$ tetraquarks from those of the $bb\bar{c}\bar{n}$ states. The main decay channel is $\bar{B}_c^*\bar{B}_s^*$ for the higher $0^+$ $bb\bar{c}\bar{s}$, while it is $\bar{B}_c\bar{B}_s$ for the lower one. The partial width ratios of the dominant channels for the highest, intermediate, and lowest $1^+$ $bb\bar{c}\bar{s}$ tetraquarks are $\Gamma_{\bar{B}_c^*\bar{B}_s^*}:\Gamma_{\bar{B}_c^*\bar{B}_s}=5.1$, $\Gamma_{\bar{B}_c^*\bar{B}_s^*}:\Gamma_{\bar{B}_c^*\bar{B}_s}:\Gamma_{\bar{B}_c\bar{B}_s^*}=1.0:6.4:5.3$, and $\Gamma_{\bar{B}_c^*\bar{B}_s}:\Gamma_{\bar{B}_c\bar{B}_s^*}=0.6$, respectively. The $2^+$ $bb\bar{c}\bar{s}$ tetraquark can decay into $\bar{B}_c^*\bar{B}_s^*$ through $S$-wave interactions, but it also has suppressed $D$-wave decay channels $\bar{B}_c^*\bar{B}_s$, $\bar{B}_c\bar{B}_s^*$, and $\bar{B}_c\bar{B}_s$.

The $bb\bar{b}\bar{n}$ tetraquarks are distributed in the range of 15,276.9$\sim$15,379.7 MeV. The two $0^+$ states have the decay channels $\Upsilon\bar{B}^*$ and $\eta_b \bar{B}$, with the former (latter) being dominant for the higher (lower) tetraquark. The $1^+$ tetraquark states have three decay channels. The highest $1^+$ state mainly decays into $\Upsilon\bar{B}^*$, while the other two mainly decay into $\Upsilon\bar{B}$ and $\eta_b\bar{B}^*$. The partial width ratios of the dominant channels for the intermediate and lowest states are $\Gamma_{\Upsilon\bar{B}}:\Gamma_{\eta_b\bar{B}^*}=3.0$ and 0.4, respectively. The $2^+$ tetraquark has one $S$-wave channel $\Upsilon\bar{B}^*$ and three suppressed $D$-wave channels.

As for the $bb\bar{b}\bar{s}$ tetraquarks, the decay channels can be obtained from the $bb\bar{b}\bar{n}$ states by replacing $\bar{B}^{(*)}$ with $\bar{B}_s^{(*)}$ in the final states. The partial width ratios of the dominant channels for the intermediate and lowest $1^+$ $bb\bar{b}\bar{s}$ tetraquarks are $\Gamma_{\Upsilon\bar{B}_s}:\Gamma_{\eta_b\bar{B}_s^*}=2.3$ and 0.5, respectively.

\begin{table}[htbp]
\caption{Numerical results for the $bb\bar{Q}\bar{q}$ systems in MeV units. The lower limits for the tetraquark masses in the seventh column are calculated using the reference meson-meson states $B_c^-\bar{B}$, $B_c^-\bar{B}^0_s$, $\Upsilon \bar{B}$, and $\Upsilon \bar{B}^0_s$ for the $bb\bar{c}\bar{n}$, $bb\bar{c}\bar{s}$, $bb\bar{b}\bar{n}$, and $bb\bar{b}\bar{s}$ systems, respectively. The tetraquark masses in the sixth column are calculated using $X(4140)$ as the reference state. The corresponding $\langle H_{CMI}\rangle$ base vectors are given in Sec. \ref{basevector}.}\label{mass of bbQq}\scriptsize
\begin{tabular}{cccccccc}\hline
System&$J^{P}$ & $\langle H_{CMI}\rangle$ &Eigenvalue &Eigenvector &Mass& Lower limit &Upper limit\\\hline
$bb\bar{c}\bar{n}$&	$2^{+}$ &$\left(\begin{array}{c}30.1\end{array}\right)$&$\left(\begin{array}{c}30.1\end{array}\right)$&$\left[\begin{array}{c}\{1.00\}\end{array}\right]$&$\left(\begin{array}{c}12020.9\end{array}\right)$&$\left(\begin{array}{c}11670.9\end{array}\right)$&$\left(\begin{array}{c}12224.8\end{array}\right)$\\
 &$1^{+}$ &$\left(\begin{array}{ccc}1.3&4.5&9.6\\4.5&-26.9&-30.5\\9.6&-30.5&2.3\end{array}\right)$&$\left(\begin{array}{c}23.1\\1.4\\-47.8\end{array}\right)$&$\left[\begin{array}{ccc}\{0.27,-0.48,0.83\}\\\{0.95,0.29,-0.13\}\\\{0.18,-0.82,-0.54\}\end{array}\right]$&$\left(\begin{array}{c}12013.9\\11992.2\\11942.9\end{array}\right)$&$\left(\begin{array}{c}11663.9\\11642.2\\11593.0\end{array}\right)$&$\left(\begin{array}{c}12217.8\\12196.1\\12146.9\end{array}\right)$\\
&$0^{+}$ &$\left(\begin{array}{cc}-13.1&52.9\\52.9&23.6\end{array}\right)$&$\left(\begin{array}{c}61.3\\-50.7\end{array}\right)$&$\left[\begin{array}{cc}\{0.58,0.81\}\\\{-0.81,0.58\}\end{array}\right]$&$\left(\begin{array}{c}12052.0\\11940.0\end{array}\right)$&$\left(\begin{array}{c}11702.1\\11590.1\end{array}\right)$&$\left(\begin{array}{c}12256.0\\12144.0\end{array}\right)$\\\hline
$bb\bar{c}\bar{s}$& $2^{+}$ &$\left(\begin{array}{c}31.5\end{array}\right)$&$\left(\begin{array}{c}31.5\end{array}\right)$&$\left[\begin{array}{c}\{1.00\}\end{array}\right]$&$\left(\begin{array}{c}12122.9\end{array}\right)$&$\left(\begin{array}{c}11762.9\end{array}\right)$&$\left(\begin{array}{c}12406.8\end{array}\right)$\\
	&$1^{+}$ &$\left(\begin{array}{ccc}1.6&3.8&8.0\\3.8&-29.3&-31.7\\8.0&-31.7&1.9\end{array}\right)$&$\left(\begin{array}{c}22.7\\1.5\\-50.1\end{array}\right)$&$\left[\begin{array}{ccc}\{0.23,-0.49,0.84\}\\\{0.96,0.24,-0.12\}\\\{0.14,-0.84,-0.53\}\end{array}\right]$&$\left(\begin{array}{c}12114.2\\12093.0\\12041.3\end{array}\right)$&$\left(\begin{array}{c}11754.1\\11732.9\\11681.3\end{array}\right)$&$\left(\begin{array}{c}12398.0\\12376.8\\12325.2\end{array}\right)$\\
&$0^{+}$ &$\left(\begin{array}{cc}-13.3&54.9\\54.9&24.8\end{array}\right)$&$\left(\begin{array}{c}63.8\\-52.4\end{array}\right)$&$\left[\begin{array}{cc}\{0.58,0.81\}\\\{-0.81,0.58\}\end{array}\right]$&$\left(\begin{array}{c}12155.3\\12039.1\end{array}\right)$&$\left(\begin{array}{c}11795.2\\11679.1\end{array}\right)$&$\left(\begin{array}{c}12439.1\\12322.9\end{array}\right)$\\
			\hline
$bb\bar{b}\bar{n}$&	$2^{+}$ &$\left(\begin{array}{c}21.9\end{array}\right)$&$\left(\begin{array}{c}21.9\end{array}\right)$&$\left[\begin{array}{c}\{1.00\}\end{array}\right]$&$\left(\begin{array}{c}15352.8\end{array}\right)$&$\left(\begin{array}{c}14779.8\end{array}\right)$&$\left(\begin{array}{c}15546.9\end{array}\right)$\\
	&$1^{+}$ &$\left(\begin{array}{ccc}-4.8&3.0&6.4\\3.0&-5.3&-28.3\\6.4&-28.3&5.9\end{array}\right)$&$\left(\begin{array}{c}29.4\\-3.5\\-30.2\end{array}\right)$&$\left[\begin{array}{ccc}\{0.09,-0.62,0.78\}\\\{0.97,0.25,0.09\}\\\{0.25,-0.74,-0.62\}\end{array}\right]$&$\left(\begin{array}{c}15360.3\\15327.5\\15300.8\end{array}\right)$&$\left(\begin{array}{c}14787.3\\14754.5\\14727.8\end{array}\right)$&$\left(\begin{array}{c}15554.4\\15521.5\\15494.8\end{array}\right)$\\
&$0^{+}$ &$\left(\begin{array}{cc}-18.1&49.0\\49.0&12.8\end{array}\right)$&$\left(\begin{array}{c}48.7\\-54.0\end{array}\right)$&$\left[\begin{array}{cc}\{-0.59,-0.81\}\\\{-0.81,0.59\}\end{array}\right]$&$\left(\begin{array}{c}15379.7\\15276.9\end{array}\right)$&$\left(\begin{array}{c}14806.6\\14703.9\end{array}\right)$&$\left(\begin{array}{c}15573.7\\15471.0\end{array}\right)$\\\hline
$bb\bar{b}\bar{s}$&	$2^{+}$ &$\left(\begin{array}{c}22.4\end{array}\right)$&$\left(\begin{array}{c}22.4\end{array}\right)$&$\left[\begin{array}{c}\{1.00\}\end{array}\right]$&$\left(\begin{array}{c}15454.1\end{array}\right)$&$\left(\begin{array}{c}14870.9\end{array}\right)$&$\left(\begin{array}{c}15728.0\end{array}\right)$\\
	&	$1^{+}$ &$\left(\begin{array}{ccc}-5.3&2.3&4.8\\2.3&-5.3&-29.4\\4.8&-29.4&5.9\end{array}\right)$&$\left(\begin{array}{c}30.4\\-4.6\\-30.6\end{array}\right)$&$\left[\begin{array}{ccc}\{0.06,-0.63,0.77\}\\\{0.98,0.18,0.07\}\\\{0.19,-0.75,-0.63\}\end{array}\right]$&$\left(\begin{array}{c}15462.0\\15427.1\\15401.1\end{array}\right)$&$\left(\begin{array}{c}14878.9\\14844.0\\14817.9\end{array}\right)$&$\left(\begin{array}{c}15736.0\\15701.0\\15675.0\end{array}\right)$\\
&$0^{+}$ &$\left(\begin{array}{cc}-19.2&50.9\\50.9&12.8\end{array}\right)$&$\left(\begin{array}{c}50.2\\-56.6\end{array}\right)$&$\left[\begin{array}{cc}\{-0.59,-0.81\}\\\{-0.81,0.59\}\end{array}\right]$&$\left(\begin{array}{c}15481.9\\15375.1\end{array}\right)$&$\left(\begin{array}{c}14898.7\\14791.9\end{array}\right)$&$\left(\begin{array}{c}15755.8\\15649.0\end{array}\right)$\\
	\hline
\end{tabular}
\end{table}

\begin{table}[htbp]
\centering
\caption{Rearrangement decays for the $bb\bar{c}\bar{n}$, $bb\bar{c}\bar{s}$, $bb\bar{b}\bar{n}$, and $bb\bar{b}\bar{s}$ states. The numbers in the parentheses are $(100|\mathcal{M}|^2/{\cal C}^2, \Gamma)$. The tetraquark mass, partial width $\Gamma$, and total width $\Gamma_{sum}$ values are given in MeV units.}\label{decay of bbQq}\scriptsize
\begin{tabular}{ccccccc}\hline\hline
System& $J^P$ & Mass & \multicolumn{3}{c}{Decay channels} & $\Gamma_{sum}$ \\ \hline
$bb\bar{c}\bar{n}$&	& & $B_c^{*-} \bar{B}^*$ & & & \\
	&$2^+$ & $\left[\begin{array}{c} 12020.9\end{array} \right]$ & $\left[\begin{array}{c}(33.3, 59.0)  \end{array}\right]$ & & & $\left[\begin{array}{c}59.0  \end{array}\right]$ \\
		&&&$B_c^{*-} \bar{B}^*$&$B_c^{*-}\bar{B}$&$B_c^-\bar{B}^*$& \\
    &$1^+$ & $\left[\begin{array}{c}12013.9\\11992.2\\11942.9 \end{array}\right]$&  $\left[\begin{array}{c}(46.1, 80.8)\\ (3.9, 6.6)\\ (0.1, 0.1) \end{array}\right]$ &  $\left[\begin{array}{c}(9.6, 17.8)\\ (17.4, 31.6)\\ (14.7, 25.0) \end{array}\right]$ &  $\left[\begin{array}{c}(0.8, 1.6)\\ (12.6, 23.7)\\ (28.2, 49.8) \end{array}\right]$ &  $\left[\begin{array}{c}100.2\\61.9\\75.0 \end{array}\right]$\\
		&&&$B_c^{*-}\bar{B}^*$&$B_c^-\bar{B}$&&\\
	&$0^+$& $\left[\begin{array}{c}12052.0\\11940.0 \end{array}\right]$& $\left[\begin{array}{c}(55.3,101.6)\\ (3.1, 4.8) \end{array}\right]$& $\left[\begin{array}{c}(0.2, 0.4)\\ (41.5, 77.7) \end{array}\right]$& & $\left[\begin{array}{c}102.0\\82.5 \end{array}\right]$\\ \hline
$bb\bar{c}\bar{s}$&			&&$B_c^{*-}\bar{B}_s^{*0}$&&&\\
	&$2^+$& $\left[\begin{array}{c}12122.9 \end{array}\right]$& $\left[\begin{array}{c}(33.3, 59.2) \end{array}\right]$&&& $\left[\begin{array}{c}59.2 \end{array}\right]$\\
		&&&$B_c^{*-}\bar{B}_s^{*0}$&$B_c^{*-}\bar{B}_s^0$&$B_c^-\bar{B}_s^{*0}$& \\
    &$1^+$& $\left[\begin{array}{c}12114.2\\12093.0\\12041.3 \end{array}\right]$& $\left[\begin{array}{c}(47.0, 82.7)\\ (2.8, 4.9)\\ (0.1, 0.2) \end{array}\right]$& $\left[\begin{array}{c}(8.6, 16.1)\\ (17.1, 31.2)\\ (16.0, 27.3) \end{array}\right]$& $\left[\begin{array}{c}(1.1, 2.2)\\ (13.9, 26.1)\\ (26.7, 47.1) \end{array}\right]$& $\left[\begin{array}{c}100.9\\62.2\\74.6 \end{array}\right]$\\
		&&&$B_c^{*-}\bar{B}_s^{*0}$&$B_c\bar{B}_s^0$&&\\
	&$0^+$& $\left[\begin{array}{c}12155.3\\12039.1 \end{array}\right]$& $\left[\begin{array}{c}(55.3, 102.0)\\ (3.0, 4.8) \end{array}\right]$& $\left[\begin{array}{c}(0.2, 0.4)\\ (41.5, 77.9) \end{array}\right]$&& $\left[\begin{array}{c}102.4\\82.7 \end{array}\right]$\\	\hline
$bb\bar{b}\bar{n}$&			&&$\Upsilon\bar{B}^*$&&&\\
	&$2^+$& $\left[\begin{array}{c}15352.8 \end{array}\right]$& $\left[\begin{array}{c}(33.3, 50.1) \end{array}\right]$&&& $\left[\begin{array}{c}50.1 \end{array}\right]$\\
		&&&$\Upsilon\bar{B}^*$&$\Upsilon\bar{B}$&$\eta_b\bar{B}^*$&\\
    &$1^+$& $\left[\begin{array}{c}15360.3\\15327.5\\15300.8 \end{array}\right]$& $\left[\begin{array}{c}(49.4, 74.7)\\ (0.3, 0.4)\\ (0.3, 0.5) \end{array}\right]$& $\left[\begin{array}{c}(3.0, 4.7)\\ (25.1, 38.4)\\ (13.6, 20.4) \end{array}\right]$& $\left[\begin{array}{c}(1.0, 1.6)\\ (8.3, 12.8)\\ (32.4, 49.4) \end{array}\right]$& $\left[\begin{array}{c}81.0\\51.6\\70.3 \end{array}\right]$\\
		&&&$\Upsilon\bar{B}^*$&$\eta_b\bar{B}$&&\\
	&$0^+$& $\left[\begin{array}{c}15379.7\\15276.9 \end{array}\right]$& $\left[\begin{array}{c}(54.9, 84.3)\\ (3.4, 4.8) \end{array}\right]$& $\left[\begin{array}{c}(0.1, 0.2)\\ (41.6, 64.6) \end{array}\right]$&& $\left[\begin{array}{c}84.5\\69.4 \end{array}\right]$\\		\hline
$bb\bar{b}\bar{s}$& 	&&$\Upsilon\bar{B}_s^{*0}$&&&\\
	&$2^+$& $\left[\begin{array}{c}15454.1 \end{array}\right]$& $\left[\begin{array}{c}(33.3, 50.2) \end{array}\right]$&&& $\left[\begin{array}{c}50.2 \end{array}\right]$\\
		&&&$\Upsilon\bar{B}_s^{*0}$&$\Upsilon\bar{B}_s^0$&$\eta_b\bar{B}_s^{*0}$&\\
    &$1^+$& $\left[\begin{array}{c}15462.0\\15427.1\\15401.1 \end{array}\right]$& $\left[\begin{array}{c}(49.5, 75.1)\\ (0.1, 0.2)\\ (0.3, 0.5) \end{array}\right]$& $\left[\begin{array}{c}(2.5, 4.0)\\ (23.3, 35.7)\\ (15.9, 23.9) \end{array}\right]$& $\left[\begin{array}{c}(1.1, 1.8)\\ (10.1, 15.7)\\ (30.4, 46.4) \end{array}\right]$& $\left[\begin{array}{c}80.8\\51.6\\70.8 \end{array}\right]$\\
		&&&$\Upsilon\bar{B}_s^{*0}$&$\eta_b\bar{B}_s^0$&&\\
	&$0^+$& $\left[\begin{array}{c}15481.9\\15375.1 \end{array}\right]$& $\left[\begin{array}{c}(54.9, 84.4)\\ (3.5, 4.9) \end{array}\right]$& $\left[\begin{array}{c}(0.1, 0.2)\\ (41.6, 64.7) \end{array}\right]$&& $\left[\begin{array}{c}84.6\\69.6 \end{array}\right]$\\
\hline\hline
\end{tabular}
\end{table}

\subsection{The $bc\bar{c}\bar{n}$, $bc\bar{c}\bar{s}$, $bc\bar{b}\bar{n}$, and $bc\bar{b}\bar{s}$ states}

The Pauli principle does not place constraints on the tetraquark wave functions, and the total number of states in each system is twelve. We present the CMI- and tetraquark-mass-related results in table \ref{mass of bcQq} and plot the mass spectra using $X(4140)$ as the reference state in Fig. \ref{triply figures}(i)-(l). When obtaining the lower limits for the tetraquark masses in Table \ref{mass of bcQq}, we adopt the meson-meson thresholds of $B_c^-D$, $B_c^-D_s^+$, $\Upsilon D$, and $\Upsilon D_s^+$ for the $bc\bar{c}\bar{n}$, $bc\bar{c}\bar{s}$, $bc\bar{b}\bar{n}$, and $bc\bar{b}\bar{s}$ systems, respectively. Similar values for the lower and upper limits can be found in Ref. \cite{Chen:2016ont}. For these $bc\bar{Q}\bar{q}$ systems, an alternative selection of reference meson-meson states is possible. When the meson-meson thresholds of $\bar{B}J/\psi$, $\bar{B}_sJ/\psi$, $\bar{B}B_c$, and $\bar{B}_sB_c$ were used to estimate the tetraquark masses in Ref. \cite{Chen:2016ont}, values larger than the lower limits were obtained, but they were smaller than the predicted values that we obtain here. Our $bc\bar{Q}\bar{q}$ tetraquark masses are smaller than the upper limits. As shown in Fig. \ref{triply figures}, more rearrangement decay channels exist than in the previous cases, and all the $bc\bar{Q}\bar{q}$ tetraquarks are unstable. Table \ref{decay of bcQq} lists the decay information in our scheme. One may calculate the partial width ratios of different channels for a given tetraquark using this table.

The $bc\bar{c}\bar{n}$ tetraquarks have masses that range from 8478.1 MeV to 8746.8 MeV. For the highest and second-highest $0^+$ states, the dominant channel partial width ratios are $\Gamma_{\bar{B}_c^*D^*}:\Gamma_{\bar{B}^*J/\psi}=1.2$ and 0.8, respectively. For the second-lowest and lowest $0^+$ states, these are $\Gamma_{\bar{B}_cD}:\Gamma_{\bar{B}\eta_c}=0.4$ and 2.3, respectively. There are six $1^+$ $bc\bar{c}\bar{n}$ tetraquarks. From the highest to lowest, the partial width ratios of their dominant channels are
\begin{equation}
	\begin{aligned}
		\Gamma_{B_c^{*-}D^*}:\Gamma_{\bar{B}^*J/\psi}:\Gamma_{\bar{B}J/\psi}&=7.1:1.6:1.0,\\
		\Gamma_{B_c^{*-}D^*}:\Gamma_{B_c^-D^*}:\Gamma_{\bar{B}^*J/\psi}&=2.3:1.0:9.4,\\
		\Gamma_{B_c^-D^*}:\Gamma_{\bar{B}^*\eta_c}:\Gamma_{\bar{B}J/\psi}&=4.1:1.0:10.5,\\
		\Gamma_{B_c^-D^*}:\Gamma_{\bar{B}J/\psi}&=1.8,\\
		\Gamma_{B_c^{*-}D}:\Gamma_{B_c^-D^*}:\Gamma_{\bar{B}^*\eta_c}:\Gamma_{\bar{B}J/\psi}&=6.5:1.9:12.4:1.0,\\
		\Gamma_{B_c^{*-}D}:\Gamma_{\bar{B}^*\eta_c}&=1.7.
	\end{aligned}
\end{equation}
For the $S$-wave decays of the two $2^+$ tetraquarks, the higher state has a partial width ratio of $\Gamma_{\bar{B}_c^*D^*}:\Gamma_{\bar{B}^*J/\psi}=4.4$, while the lower one mainly decays into $\bar{B}^*J/\psi$.

The twelve $bc\bar{c}\bar{s}$ tetraquark masses are about 90 MeV heavier than those of the $bc\bar{c}\bar{n}$ states. We easily obtain their $S$-wave decay channels from the $bc\bar{c}\bar{n}$ states by replacing an $n$ quark with an $s$ quark in the final states. The rearrangement decay properties of these two systems have similar features.

The $bc\bar{b}\bar{n}$ tetraquarks are located in the 11,829.5$\sim$12,059.3 MeV mass range and the $bc\bar{b}\bar{s}$ states are about 100 MeV heavier. These two systems' rearrangement decays are also similar.

\setlength{\tabcolsep}{0.1mm}
\begin{table}[htbp]
\caption{Numerical results for the $bc\bar{Q}\bar{q}$ systems in MeV units. The tetraquark mass lower limits in the seventh column are calculated using the reference meson-meson states $B_c^-D$, $B_c^-D_s^+$, $\Upsilon D$, and $\Upsilon D_s^+$ for the $bc\bar{c}\bar{n}$, $bc\bar{c}\bar{s}$, $bc\bar{b}\bar{n}$, and $bc\bar{b}\bar{s}$ systems, respectively. The tetraquark masses in the sixth column are calculated using $X(4140)$ as the reference state. The corresponding $\langle H_{CMI}\rangle$ base vectors are given in Sec. \ref{basevector}.}\label{mass of bcQq}
\resizebox{\linewidth}{!}{
\begin{tabular}{cccccccc}\hline
System & $J^{P}$ & $\langle H_{CMI}\rangle$ &Eigenvalue &Eigenvector &Mass&Lower limit&Upper limit\\\hline
$bc\bar{c}\bar{n}$ &$2^{+}$ &$\left(\begin{array}{cc}49.7&-7.1\\-7.1&39.1\end{array}\right)$&$\left(\begin{array}{c}53.2\\35.5\end{array}\right)$&$\left[\begin{array}{cc}\{-0.89,0.45\}\\\{-0.45,-0.89\}\end{array}\right]$&$\left(\begin{array}{c}8713.9\\8696.2\end{array}\right)$&$\left(\begin{array}{c}8353.7\\8336.0\end{array}\right)$&$\left(\begin{array}{c}8917.6\\8899.9\end{array}\right)$\\
	& $1^{+}$ &$\left(\begin{array}{cccccc}-65.7&-0.47&30.64&7.07&26.00&-0.40\\-0.47&13.33&8.33&26.00&0.00&-48.93\\30.64&8.33&2.67&-0.40&-48.93&0.00\\7.07&26.00&-0.40&-7.07&-0.19&12.26\\26.00&0.00&-48.93&-0.19&-26.67&3.33\\-0.40&-48.93&0.00&12.26&3.33&-5.33\end{array}\right)$&$\left(\begin{array}{c}58.8\\37.8\\4.6\\-26.7\\-58.4\\-104.8\end{array}\right)$&$\left[\begin{array}{cccccc}\{0.04,0.75,0.28,0.20,-0.17,-0.54\}\\\{0.09,-0.24,0.77,-0.07,-0.54,0.21\}\\\{0.14,0.12,0.04,0.85,0.10,0.48\}\\\{0.67,-0.03,0.35,-0.16,0.63,-0.05\}\\\{-0.12,0.60,0.03,-0.44,0.07,0.65\}\\\{0.71,0.08,-0.45,-0.09,-0.52,0.07\}\end{array}\right]$&$\left(\begin{array}{c}8719.4\\8698.4\\8665.3\\8633.9\\8602.3\\8555.9\end{array}\right)$&$\left(\begin{array}{c}8359.3\\8338.3\\8305.2\\8273.8\\8242.1\\8195.7\end{array}\right)$&$\left(\begin{array}{c}8923.2\\8902.2\\8869.0\\8837.7\\8806.0\\8759.6\end{array}\right)$\\
&$0^{+}$ &$\left(\begin{array}{cccc}-30.13&-5.77&14.14&84.75\\-5.77&-48.00&84.75&0.00\\14.14&84.75&-123.33&-14.43\\84.75&0.00&-14.43&24.0\end{array}\right)$&$\left(\begin{array}{c}86.2\\7.3\\-88.4\\-182.5\end{array}\right)$&$\left[\begin{array}{cccc}\{0.59,-0.05,-0.04,0.81\}\\\{0.09,0.83,0.55,0.01\}\\\{0.79,-0.17,0.14,-0.58\}\\\{-0.17,-0.53,0.82,0.13\}\end{array}\right]$&$\left(\begin{array}{c}8746.8\\8668.0\\8572.2\\8478.1\end{array}\right)$&$\left(\begin{array}{c}8386.7\\8307.8\\8212.1\\8118.0\end{array}\right)$&$\left(\begin{array}{c}8950.6\\8871.7\\8776.0\\8681.9\end{array}\right)$\\
			\hline
$bc\bar{c}\bar{s}$&			$2^{+}$ &$\left(\begin{array}{cc}50.27&-6.79\\-6.79&40.27\end{array}\right)$&$\left(\begin{array}{c}53.7\\36.8\end{array}\right)$&$\left[\begin{array}{cc}\{-0.89,0.45\}\\\{-0.45,-0.89\}\end{array}\right]$&$\left(\begin{array}{c}8805.0\\8788.1\end{array}\right)$&$\left(\begin{array}{c}8456.9\\8440.0\end{array}\right)$&$\left(\begin{array}{c}9098.7\\9081.8\end{array}\right)$\\
 &	$1^{+}$ &$\left(\begin{array}{cccccc}-67.07&-1.89&30.17&6.79&25.60&-1.60\\-1.89&14.53&8.00&25.60&0.00&-49.78\\30.17&8.00&2.27&-1.60&-49.78&0.00\\6.79&25.60&-1.60&-6.67&-0.75&12.07\\25.60&0.00&-49.78&-0.75&29.07&3.20\\-1.60&-49.78&0.00&12.07&3.20&-4.53\end{array}\right)$&$\left(\begin{array}{c}60.0\\37.9\\4.3\\-28.9\\-57.5\\-106.4\end{array}\right)$&$\left[\begin{array}{cccccc}\{0.03,0.76,0.25,0.19,-0.15,-0.56\}\\\{0.09,-0.21,0.79,-0.08,-0.54,0.18\}\\\{0.10,0.13,0.03,0.86,0.05,0.47\}\\\{0.68,-0.03,0.35,-0.10,0.63,-0.05\}\\\{-0.11,0.60,0.05,-0.44,0.10,0.65\}\\\{0.71,0.10,-0.45,-0.09,-0.53,0.09\}\end{array}\right]$&$\left(\begin{array}{c}8811.3\\8789.2\\8755.6\\8722.4\\8693.8\\8644.8\end{array}\right)$&$\left(\begin{array}{c}8463.2\\8441.1\\8407.6\\8374.3\\8345.7\\8296.8\end{array}\right)$&$\left(\begin{array}{c}9105.0\\9082.9\\9049.3\\9016.1\\8987.5\\8938.6\end{array}\right)$\\
&$0^{+}$ &$\left(\begin{array}{cccc}-30.13&-5.54&13.58&86.22\\-5.54&-50.40&86.22&0.00\\13.58&86.22&-125.73&13.86\\86.22&0.00&-13.86&25.20\end{array}\right)$&$\left(\begin{array}{c}88.3\\6.2\\-89.8\\-185.8\end{array}\right)$&$\left[\begin{array}{cccc}\{0.59,-0.04,-0.03,0.81\}\\\{0.09,0.83,0.55,0.00\}\\\{0.79,-0.16,0.13,-0.58\}\\\{-0.16,-0.53,0.82,0.12\}\end{array}\right]$&$\left(\begin{array}{c}8839.6\\8757.5\\8661.5\\8565.4\end{array}\right)$&$\left(\begin{array}{c}8491.5\\8409.5\\8313.4\\8217.4\end{array}\right)$&$\left(\begin{array}{c}9133.3\\9051.2\\8955.2\\8859.2\end{array}\right)$\\
			\hline
$bc\bar{b}\bar{n}$&	$2^{+}$ &$\left(\begin{array}{cc}45.27&-11.60\\-11.60&28.67\end{array}\right)$&$\left(\begin{array}{c}51.2\\22.7\end{array}\right)$&$\left[\begin{array}{cc}\{-0.89,0.46\}\\\{-0.46,-0.89\}\end{array}\right]$&$\left(\begin{array}{c}12042.0\\12013.5\end{array}\right)$&$\left(\begin{array}{c}11468.9\\11440.4\end{array}\right)$&$\left(\begin{array}{c}12245.9\\12217.4\end{array}\right)$\\
  &		$1^{+}$ &$\left(\begin{array}{cccccc}-54.07&-11.79&23.10&11.60&19.60&-10.00\\-11.79&2.53&13.67&19.60&0.00&-42.14\\23.10&13.67&6.27&-10.00&-42.14&0.00\\11.60&19.60&-10.00&-11.07&-4.71&9.24\\19.60&0.00&-42.14&-4.71&-5.07&5.47\\-10.00&-42.14&0.00&9.24&5.47&-12.53\end{array}\right)$&$\left(\begin{array}{c}51.4\\32.3\\-3.5\\-17.5\\-43.3\\-93.3\end{array}\right)$&$\left[\begin{array}{cccccc}\{0.03,0.50,0.61,0.05,-0.49,-0.37\}\\\{-0.03,0.60,-0.46,0.23,0.41,-0.46\}\\\{0.02,0.08,-0.03,0.89,-0.18,0.41\}\\\{0.66,-0.06,0.46,0.12,0.58,0.00\}\\\{-0.35,0.52,0.23,-0.25,0.32,0.62\}\\\{0.67,0.33,-0.38,-0.27,-0.36,0.31\}\end{array}\right]$&$\left(\begin{array}{c}12042.1\\12023.1\\11987.3\\11973.2\\11947.5\\11897.5\end{array}\right)$&$\left(\begin{array}{c}11469.0\\11450.0\\11414.2\\11400.1\\11374.3\\11324.4\end{array}\right)$&$\left(\begin{array}{c}12246.1\\12227.0\\12191.2\\12177.2\\12151.4\\12101.4\end{array}\right)$\\
&$0^{+}$ &$\left(\begin{array}{cccc}-30.93&-9.47&23.19&73.00\\-9.47&-26.40&72.99&0.00\\23.19&72.99&-103.73&-23.67\\72.99&0.00&-23.67&13.2\end{array}\right)$&$\left(\begin{array}{c}68.5\\17.5\\-72.7\\-161.2\end{array}\right)$&$\left[\begin{array}{cccc}\{0.58,-0.12,-0.08,0.80\}\\\{0.17,0.83,0.52,0.06\}\\\{0.73,-0.29,0.28,-0.55\}\\\{-0.31,-0.46,0.80,0.24\}\end{array}\right]$&$\left(\begin{array}{c}12059.3\\12008.3\\11918.1\\11829.5\end{array}\right)$&$\left(\begin{array}{c}11486.2\\11435.2\\11345.0\\11256.4\end{array}\right)$&$\left(\begin{array}{c}12263.2\\12212.2\\12122.0\\12033.5\end{array}\right)$\\
			\hline
$bc\bar{b}\bar{s}$&			$2^{+}$ &$\left(\begin{array}{cc}46.27&-11.31\\-11.31&29.07\end{array}\right)$&$\left(\begin{array}{c}51.9\\23.5\end{array}\right)$&$\left[\begin{array}{cc}\{-0.90,0.44\}\\\{-0.44,-0.90\}\end{array}\right]$&$\left(\begin{array}{c}12143.3\\12114.9\end{array}\right)$&$\left(\begin{array}{c}11572.2\\11543.8\end{array}\right)$&$\left(\begin{array}{c}12427.2\\12398.8\end{array}\right)$\\
  &		$1^{+}$ &$\left(\begin{array}{cccccc}-55.07&-13.20&22.63&11.31&19.20&-11.20\\-13.20&2.53&13.33&19.20&0.00&-42.99\\22.63&13.33&6.27&-11.20&-42.99&0.00\\11.31&19.20&-11.20&-11.47&-5.28&9.05\\19.20&0.00&-42.99&-5.28&-5.07&5.33\\-11.20&-42.99&0.00&9.05&5.33&-12.53\end{array}\right)$&$\left(\begin{array}{c}51.9\\33.4\\-3.7\\-19.2\\-42.7\\-94.9\end{array}\right)$&$\left[\begin{array}{cccccc}\{0.02,0.50,0.61,0.03,-0.49,-0.37\}\\\{-0.03,0.60,-0.46,0.22,0.40,-0.46\}\\\{-0.03,0.08,-0.07,0.88,-0.23,0.39\}\\\{0.67,-0.06,0.45,0.21,0.55,0.00\}\\\{-0.33,0.52,0.26,-0.23,0.34,0.62\}\\\{0.66,0.34,-0.37,-0.28,-0.36,0.32\}\end{array}\right]$&$\left(\begin{array}{c}12143.3\\12124.9\\12087.7\\12072.2\\12048.7\\11996.6\end{array}\right)$&$\left(\begin{array}{c}11572.2\\11553.8\\11516.6\\11501.1\\11477.6\\11425.4\end{array}\right)$&$\left(\begin{array}{c}12427.2\\12408.7\\12371.6\\12356.1\\12332.6\\12280.4\end{array}\right)$\\
&   $0^{+}$ &$\left(\begin{array}{cccc}-31.73&-9.24&22.63&74.46\\-9.24&-26.40&74.46&0.00\\22.63&74.46&-105.73&-23.09\\74.46&0.00&-23.09&13.20\end{array}\right)$&$\left(\begin{array}{c}69.6\\18.3\\-75.2\\-163.3\end{array}\right)$&$\left[\begin{array}{cccc}\{0.58,-0.12,-0.08,0.80\}\\\{0.17,0.84,0.52,0.06\}\\\{0.74,-0.28,0.28,-0.55\}\\\{-0.30,-0.46,0.80,0.23\}\end{array}\right]$&$\left(\begin{array}{c}12161.0\\12109.7\\12016.3\\11928.2\end{array}\right)$&$\left(\begin{array}{c}11589.9\\11538.6\\11445.1\\11357.0\end{array}\right)$&$\left(\begin{array}{c}12444.9\\12393.6\\12300.1\\12212.0\end{array}\right)$\\
\hline
\end{tabular}}
\end{table}

\begin{table}[htbp]
\centering
\caption{Rearrangement decays for the $bc\bar{c}\bar{n}$, $bc\bar{c}\bar{s}$, $bc\bar{b}\bar{n}$, and $bc\bar{b}\bar{s}$ states. The numbers in the parentheses are $(100|\mathcal{M}|^2/{\cal C}^2, \Gamma)$. The tetraquark mass, partial width $\Gamma$, and total width $\Gamma_{sum}$ values are given in MeV units.}\label{decay of bcQq}\scriptsize
\begin{tabular}{cccccccccc}\hline\hline
System& $J^P$&Mass&\multicolumn{6}{c}{Decay channels}&$\Gamma_{sum}$\\
			\hline
$bc\bar{c}\bar{n}$&			&&$B_c^{*-}D^*$&$\bar{B}^*J/\psi$&&&&&\\
  &			$2^+$&$\left[\begin{array}{c}8713.9\\8696.2 \end{array}\right]$&$\left[\begin{array}{c}(97.3, 245.8)\\ (2.1, 5.3) \end{array}\right]$&$\left[\begin{array}{c}(21.8, 55.2)\\ (77.7, 191.3) \end{array}\right]$&&&&&$\left[\begin{array}{c}301.1\\196.6 \end{array}\right]$\\
			&&&$B_c^{*-}D^*$&$B_c^{*-}D$&$B_c^-D^*$&$\bar{B}^*J/\psi$&$\bar{B}^*\eta_c$&$\bar{B}J/\psi$&\\
	&		$1^+$&$\left[\begin{array}{c}8719.4\\8698.4\\8665.3\\8633.9\\8602.3\\8555.9 \end{array}\right]$&$\left[\begin{array}{c}(78.2, 199.0)\\ (19.4, 48.1)\\ (2.1, 4.9)\\ (0.3, 0.6)\\ (0.5, 1.0)\\ (0.1, 0.2) \end{array}\right]$&$\left[\begin{array}{c}(0.1, 0.2)\\ (1.3, 3.8)\\ (1.5, 4.4)\\ (1.0, 2.8)\\ (26.1, 68.7)\\ (69.5, 173.3) \end{array}\right]$&$\left[\begin{array}{c}(2.1, 5.7)\\ (7.6, 20.8)\\ (16.7, 43.9)\\ (64.7, 163.6)\\ (8.4, 20.4)\\ (0.4, 0.8) \end{array}\right]$&$\left[\begin{array}{c}(17.8, 45.6)\\ (79.1, 195.6)\\ (1.2, 2.8)\\ (0.3, 0.8)\\ (0.9, 1.7)\\ (0.4, 0.8) \end{array}\right]$&$\left[\begin{array}{c}(1.2, 3.7)\\ (1.0, 2.8)\\ (3.8, 10.6)\\ (0.1, 0.2)\\ (51.0, 131.6)\\ (42.8, 102.1) \end{array}\right]$&$\left[\begin{array}{c}(10.3, 28.2)\\ (2.1, 5.6)\\ (43.6, 111.1)\\ (38.4, 92.8)\\ (4.8, 10.6)\\ (0.9, 1.7) \end{array}\right]$&$\left[\begin{array}{c}282.3\\276.7\\177.7\\260.6\\233.9\\278.9 \end{array}\right]$\\
			&&&$B_c^{*-}D^*$&$B_c^-D$&$\bar{B}^*J/\psi$&$\bar{B}\eta_c$&&&\\
	&$0^+$&$\left[\begin{array}{c}8746.8\\8668.0\\8572.2\\8478.1 \end{array}\right]$&$\left[\begin{array}{c}(61.5, 161.5)\\ (36.8, 87.5)\\ (2.4, 4.8)\\ (0.1, 0.1) \end{array}\right]$&$\left[\begin{array}{c}(0.1, 0.2)\\ (1.2, 3.6)\\ (23.4, 65.3)\\ (75.9, 190.2) \end{array}\right]$&$\left[\begin{array}{c}(49.3, 130.8)\\ (45.2, 106.1)\\ (4.4, 8.2)\\ (1.3, 1.5) \end{array}\right]$&$\left[\begin{array}{c}(0.3, 0.8)\\ (3.6, 10.8)\\ (60.8, 161.9)\\ (35.4, 80.0) \end{array}\right]$&&&$\left[\begin{array}{c}293.2\\208.0\\240.2\\271.8 \end{array}\right]$\\
			\hline
$bc\bar{c}\bar{s}$& 		&&$B_c^{*-}D_s^{*+}$&$\bar{B}_s^{*0}J/\psi$&&&&&\\
	&$2^+$&$\left[\begin{array}{c}8805.0\\8788.1\end{array}\right]$&$\left[\begin{array}{c}(97.3, 240.1)\\ (2.1, 5.2)\end{array}\right]$&$\left[\begin{array}{c}(21.8, 54.3)\\ (77.7, 188.5) \end{array}\right]$&&&&&$\left[\begin{array}{c}294.3\\193.7 \end{array}\right]$\\
			&&&$B_c^{*-}D_s^{*+}$&$B_c^{*-}D_s^+$&$B_c^-D_s^{*+}$&$\bar{B}_s^{*0}J/\psi$&$\bar{B}_s^{*0}\eta_c$&$\bar{B}_s^0J/\psi$&\\
    &$1^+$&$\left[\begin{array}{c}8811.3\\8789.2\\8755.6\\8722.4\\8693.8\\8644.8 \end{array}\right]$&$\left[\begin{array}{c}(76.2, 189.5)\\ (23.2, 56.1)\\ (1.4, 3.3)\\ (0.3, 0.6)\\ (0.5, 1.0)\\ (0.1, 0.1) \end{array}\right]$&$\left[\begin{array}{c}(0.1, 0.4)\\ (1.3, 3.8)\\ (1.7, 4.8)\\ (0.7, 1.8)\\ (24.9, 64.6)\\ (72.3, 176.0) \end{array}\right]$&$\left[\begin{array}{c}(2.3, 6.2)\\ (8.1, 21.7)\\ (20.7, 53.2)\\ (61.6, 151.7)\\ (7.1, 16.9)\\ (0.2, 0.4) \end{array}\right]$&$\left[\begin{array}{c}(21.3, 53.7)\\ (75.9, 184.6)\\ (1.3, 3.0)\\ (0.3, 0.7)\\ (0.9, 1.7)\\ (0.4, 0.7) \end{array}\right]$&$\left[\begin{array}{c}(1.3, 3.7)\\ (1.3, 3.6)\\ (3.7, 10.4)\\ (0.3, 0.8)\\ (52.6, 133.4)\\ (41.4, 96.7) \end{array}\right]$&$\left[\begin{array}{c}(9.1, 24.8)\\ (2.3, 6.1)\\ (39.0, 98.0)\\ (42.2, 100.6)\\ (5.7, 12.8)\\ (1.1, 2.1) \end{array}\right]$&$\left[\begin{array}{c}278.1\\276.0\\172.6\\256.1\\230.4\\276.1 \end{array}\right]$\\
			&&&$B_c^{*-}D_s^{*+}$&$B_c^-D_s^+$&$\bar{B}_s^{*0}J/\psi$&$\bar{B}_s^0\eta_c$&&&\\
 	 &$0^+$&$\left[\begin{array}{c}8839.6\\8757.5\\8661.5\\8565.4 \end{array}\right]$&$\left[\begin{array}{c}(60.1, 154.6)\\ (37.7, 87.0)\\ (2.4, 4.7)\\ (0.1, 0.1) \end{array}\right]$&$\left[\begin{array}{c}(0.1, 0.2)\\ (1.1, 3.3)\\ (24.4, 66.7)\\ (74.3, 181.2) \end{array}\right]$&$\left[\begin{array}{c}(50.5, 132.3)\\ (44.3, 102.0)\\ (4.4, 8.0)\\ (1.2, 1.3) \end{array}\right]$&$\left[\begin{array}{c}(0.2, 0.7)\\ (3.8, 11.1)\\ (59.3, 155.5)\\ (36.5, 81.0) \end{array}\right]$&&&$\left[\begin{array}{c}287.8\\203.3\\234.8\\263.7 \end{array}\right]$\\\hline
$bc\bar{b}\bar{n}$& 	&&$\Upsilon D^*$&$\bar{B}^*B_c^{*+}$&&&&&\\
	&$2^+$&$\left[\begin{array}{c}12042.0\\12013.5 \end{array}\right]$&$\left[\begin{array}{c}(98.5, 174.4)\\ (1.9, 3.3) \end{array}\right]$&$\left[\begin{array}{c}(21.3, 38.6)\\ (79.1, 138.6) \end{array}\right]$&&&&&$\left[\begin{array}{c}213.0\\141.9 \end{array}\right]$\\
			&&&$\Upsilon D^*$&$\Upsilon D$&$\eta_bD^*$&$\bar{B}^*B_c^{*+}$&$\bar{B}^*B_c^+$&$\bar{B}B_c^{*+}$&\\
   &$1^+$&$\left[\begin{array}{c}12042.1\\12023.1\\11987.3\\11973.2\\11947.5\\11897.5 \end{array}\right]$&$\left[\begin{array}{c}(98.4, 174.3)\\ (1.0, 1.8)\\ (0.4, 0.7)\\ (0.0, 0.0)\\ (0.2, 0.4)\\ (0.0, 0.0) \end{array}\right]$&$\left[\begin{array}{c}(0.0, 0.0)\\ (0.5, 1.0)\\ (1.9, 3.5)\\ (0.2, 0.4)\\ (1.1, 2.0)\\ (96.1, 169.5) \end{array}\right]$&$\left[\begin{array}{c}(0.0, 0.0)\\ (8.6, 15.8)\\ (32.1, 57.9)\\ (50.7, 90.2)\\ (9.3, 16.2)\\ (0.0, 0.0) \end{array}\right]$&$\left[\begin{array}{c}(1.3, 2.3)\\ (93.5, 165.9)\\ (3.2, 5.3)\\ (0.4, 0.7)\\ (0.2, 0.3)\\ (1.9, 2.7) \end{array}\right]$&$\left[\begin{array}{c}(2.8, 5.5)\\ (0.1, 0.2)\\ (8.3, 15.5)\\ (1.0, 1.8)\\ (77.7, 138.2)\\ (9.7, 16.1) \end{array}\right]$&$\left[\begin{array}{c}(5.9, 11.3)\\ (1.4, 2.7)\\ (21.3, 38.5)\\ (57.9, 102.7)\\ (7.5, 12.8)\\ (5.8, 9.3) \end{array}\right]$&$\left[\begin{array}{c}193.5\\187.4\\121.4\\195.8\\169.9\\197.6 \end{array}\right]$\\
			&&&$\Upsilon D^*$&$\eta_bD$&$\bar{B}^*B_c^{*+}$&$\bar{B}B_c^+$&&&\\
   &$0^+$&$\left[\begin{array}{c}12059.3\\12008.3\\11918.1\\11829.5 \end{array}\right]$&$\left[\begin{array}{c}(68.2, 122.4)\\ (28.7, 49.5)\\ (2.2, 3.5)\\ (0.0, 0.0) \end{array}\right]$&$\left[\begin{array}{c}(0.0, 0.0)\\ (0.5, 0.9)\\ (9.5, 17.9)\\ (90.5, 160.6) \end{array}\right]$&$\left[\begin{array}{c}(41.0, 76.0)\\ (51.7, 89.9)\\ (4.4, 6.6)\\ (3.1, 3.8) \end{array}\right]$&$\left[\begin{array}{c}(0.3, 0.7)\\ (3.6, 7.2)\\ (75.9, 138.5)\\ (19.9, 31.9) \end{array}\right]$&&&$\left[\begin{array}{c}199.2\\147.5\\166.4\\196.3 \end{array}\right]$\\		\hline
$bc\bar{b}\bar{s}$& 	&&$\Upsilon D_s^{*+}$&$\bar{B}_s^{*0}B_c^{*+}$&&&&&\\
	&$2^+$&$\left[\begin{array}{c}12143.3\\12114.9 \end{array}\right]$&$\left[\begin{array}{c}(97.8, 172.9)\\ (2.6, 4.4) \end{array}\right]$&$\left[\begin{array}{c}(23.1, 42.1)\\ (77.2, 135.9) \end{array}\right]$&&&&&$\left[\begin{array}{c}215.0\\140.3 \end{array}\right]$\\
			&&&$\Upsilon D_s^{*+}$&$\Upsilon D_s^+$&$\eta_b D_s^{*+}$&$\bar{B}_s^{*0}B_c^{*+}$&$\bar{B}_s^{*0}B_c^+$&$\bar{B}_s^0B_c^{*+}$&\\
    &$1^+$&$\left[\begin{array}{c}12143.3\\12124.9\\12087.7\\12072.2\\12048.7\\11996.6 \end{array}\right]$&$\left[\begin{array}{c}(98.4, 173.9)\\ (1.1, 1.9)\\ (0.4, 0.6)\\ (0.0, 0.0)\\ (0.3, 0.6)\\ (0.0, 0.0) \end{array}\right]$&$\left[\begin{array}{c}(0.0, 0.0)\\ (0.6, 1.2)\\ (1.9, 3.5)\\ (0.4, 0.8)\\ (0.8, 1.5)\\ (96.3, 169.7) \end{array}\right]$&$\left[\begin{array}{c}(0.0, 0.0)\\ (8.5, 15.7)\\ (38.0, 68.3)\\ (44.6, 79.1)\\ (8.0, 14.0)\\ (0.0, 0,0) \end{array}\right]$&$\left[\begin{array}{c}(1.3, 2.3)\\ (92.8, 165.2)\\ (2.8, 4.7)\\ (0.5, 0.8)\\ (0.1, 0.2)\\ (1.8, 2.6) \end{array}\right]$&$\left[\begin{array}{c}(3.3, 6.5)\\ (0.1, 0.2)\\ (8.5, 15.9)\\ (2.4, 4.4)\\ (77.4, 138.1)\\ (8.4, 14.0) \end{array}\right]$&$\left[\begin{array}{c}(5.2, 10.1)\\ (1.3, 2.4)\\ (15.4, 27.9)\\ (62.5, 111.4)\\ (9.7, 16.7)\\ (5.9, 9.5) \end{array}\right]$&$\left[\begin{array}{c}192.9\\186.6\\120.9\\196.5\\171.1\\195.4 \end{array}\right]$\\
			&&&$\Upsilon D_s^{*+}$&$\eta_bD_s^+$&$\bar{B}_s^{*0}B_c^{*+}$&$\bar{B}_s^0B_c^+$&&&\\
	& $0^+$&$\left[\begin{array}{c}12161.0\\12109.7\\12016.3\\11928.2 \end{array}\right]$&$\left[\begin{array}{c}(68.2, 122.2)\\ (29.3, 50.3)\\ (2.2, 3.5)\\ (0.0, 0,1) \end{array}\right]$&$\left[\begin{array}{c}(0.0, 0.0)\\ (0.4, 0.8)\\ (10.0, 18.8)\\ (88.8, 157.2) \end{array}\right]$&$\left[\begin{array}{c}(41.0, 76.2)\\ (52.4, 91.5)\\ (4.0, 6.1)\\ (3.0, 3.7) \end{array}\right]$&$\left[\begin{array}{c}(0.3, 0.7)\\ (3.5, 7.0)\\ (76.3, 139.5)\\ (20.7, 33.5) \end{array}\right]$&&&$\left[\begin{array}{c}199.2\\149.7\\168.0\\194.5 \end{array}\right]$\\\hline\hline
\end{tabular}
\end{table}

\section{Discussions and summary}

Similar to fully heavy tetraquark states, a triply heavy tetraquark's exotic nature is easy to identify. In this work, we studied the spectra of triply heavy tetraquark states in a modified CMI model using a diquark-antidiquark base, with the assumption that  $X(4140)$ is the lowest $1^{++}$ compact $cs\bar{c}\bar{s}$ tetraquark state. Two-body strong decays are also studied in a simple rearrangement decay scheme. We temporarily estimated the widths by adopting the decay parameter extracted from the width of $X(6600)$, which is considered a fully charmed compact tetraquark. 

The 12 considered systems involve 96 states in total. Some tetraquarks have very similar masses, posing challenges in distinguishing them based solely on the spectrum. Fortunately, their dominant two-body decay channels and the corresponding partial width ratios are different. Therefore, the structures of exotic states could be identified by their measured masses, quantum numbers, and/or strong decay properties.

For calculations with the original CMI model, 14 coupling parameters and 4 quark masses that were extracted from the conventional hadrons are adopted. However, these parameters may introduce considerable uncertainty to tetraquark masses because of differences in the inner structures and interactions between conventional hadrons and compact tetraquark states. Since the effective quark masses are much larger than those of the effective coupling constants, the quark masses predominate this uncertainty. To reduce the uncertainty, we introduced modified mass formulae using hadron-hadron states and multiquark candidates as references \cite{Cheng:2020nho,Wu:2018xdi,Chen:2016ont}. With these methods, we obtained lower and more reasonable numerical results than the original CMI model. However, we have not yet considered the uncertainty from coupling parameters $C_{ij}$. They are hard to derive from fundamental theories. Their extraction from tetraquark states is also impossible due to the lack of awareness of exotic hadrons. The coupling parameter uncertainties are difficult to reduce right now, and, thus, should be studied in future works.

To study the dominant two-body strong decays of triply heavy tetraquark states, we introduced a simple rearrangement scheme by assuming that the Hamiltonian governing the decay processes was a constant parameter. Since there has been no observed triply heavy tetraquark candidate yet, we temporarily used the parameter extracted from the width of the possible fully charmed tetraquark $X(6600)$ and estimated each channel's partial decay width for all the studied tetraquark states. Note that each system should have a unique decay parameter, and thus, the adopted assumption is very crude. The numerical results that we present may be very different from the real values. However, the different channels' partial width ratios may rarely be dependent on this parameter, and offer valuable information to understand exotic state structures. 

In Sec. \ref{massformu}, we consider the tetraquark masses estimated with Eq. \eqref{Mass} as theoretical upper limits. Here, we introduce another method to constrain the upper limits. Suppose that there is a triply heavy hexaquark state $QQq\bar{Q}\bar{q}\bar{q}$; we can estimate its mass by using Eq. \eqref{massref} with two reference hadron-hadron states $QQ\bar{Q}\bar{q}+q\bar{q}$ and $QQq+\bar{Q}\bar{q}\bar{q}$. Previous experience \cite{Wu:2018xdi,Wu:2016gas,Li:2023wxm,Wu:2017weo} suggests that the estimated mass of a multiquark state with a reference system containing one heavy hadron and one light hadron is lighter than that with a reference system consisting of two heavy hadrons; thus, one obtains the hexaquark mass relation $M_1<M_2$, where
\begin{equation}
\begin{split}
M_1&=[M_{QQ\bar{Q}\bar{q}}-(E_{CMI})_{QQ\bar{Q}\bar{q}}]+[M_{q\bar{q}}-(E_{CMI})_{q\bar{q}}]+(E_{CMI})_{QQq\bar{Q}\bar{q}\bar{q}},\\
M_2&=[M_{QQq}-(E_{CMI})_{QQq}]+[M_{\bar{Q}\bar{q}\bar{q}}-(E_{CMI})_{\bar{Q}\bar{q}\bar{q}}]+(E_{CMI})_{QQq\bar{Q}\bar{q}\bar{q}}.
\end{split}
\end{equation}
Setting $M_1=M_2$ may constrain the upper limit for the $QQ\bar{Q}\bar{q}$ mass. Here, we just consider the case $QQq\to\Xi_{cc}$ because it is currently the only observed doubly heavy baryon. Then, the formula to constrain the upper limit reads as
\begin{eqnarray}\label{newupper}
M^{upper}_{cc\bar{Q}\bar{q}}
&=&[M_{\Xi_{cc}}-(E_{CMI})_{\Xi_{cc}}]-[M_{n\bar{q}'}-(E_{CMI})_{n\bar{q}'}]\nonumber\\
&&+[M_{\bar{Q}\bar{q}\bar{q}'}-(E_{CMI})_{\bar{Q}\bar{q}\bar{q}'}]+(E_{CMI})_{cc\bar{Q}\bar{q}}.
\end{eqnarray}
Considering different meson and baryon states, we obtained the minimum values for $M^{upper}_{cc\bar{Q}\bar{q}}$, which are collected in table \ref{newuppernum}. Here, we choose to use $K$ for the light meson and $\Xi_c$, $\Omega_c$, $\Xi_b$, and $\Omega_b$ for the baryons in the $cc\bar{c}\bar{n}$, $cc\bar{c}\bar{s}$, $cc\bar{b}\bar{n}$, and $cc\bar{b}\bar{s}$ cases, respectively. Compared with table \ref{mass of ccQq}, the values in table \ref{newuppernum} are smaller than the upper limits, but higher than the masses predicted using $X(4140)$ as the reference. Therefore, the updated upper limits are reasonable and the measured $cc\bar{Q}\bar{q}$ tetraquarks beyond these constraints should be interpreted as excited states. Note that it would be possible to obtain new constraints from the $\Omega_{cc}$ mass if it were observed.
	
\begin{table}[htbp]
\caption{Upper limits for the masses of $cc\bar{Q}\bar{q}$ states estimated with Eq. \eqref{newupper} in MeV units.}\label{newuppernum}
\begin{tabular}{c|cccc}
\hline\diagbox{$J^P$}{System}&$cc\bar{c}\bar{n}$&$cc\bar{c}\bar{s}$&$cc\bar{b}\bar{n}$&$cc\bar{b}\bar{s}$\\\hline$2^+$&$\left(\begin{array}{c}5477.7\end{array}\right)$&$\left(\begin{array}{c}5638.8\end{array}\right)$&$\left(\begin{array}{c}8793.2\end{array}\right)$&$\left(\begin{array}{c}8944.1\end{array}\right)$\\$1^+$&$\left(\begin{array}{c}5488.2\\5415.5\\5348.1\end{array}\right)$&$\left(\begin{array}{c}5647.7\\5576.2\\5505.8\end{array}\right)$&$\left(\begin{array}{c}8817.6\\8752.8\\8688.8\end{array}\right)$&$\left(\begin{array}{c}8968.9\\8903.7\\8838.2\end{array}\right)$\\$0^+$&$\left(\begin{array}{c}5541.3\\5296.9\end{array}\right)$&$\left(\begin{array}{c}5703.1\\5456.5\end{array}\right)$&$\left(\begin{array}{c}8844.8\\8642.0\end{array}\right)$&$\left(\begin{array}{c}8996.1\\8791.4\end{array}\right)$\\\hline
\end{tabular}
\end{table}
	
In the above discussions, we introduced a reference system to reduce the mass uncertainty for tetraquark states from effective quark masses. The uncertainties are now governed by those of the quark mass gaps $\Delta_{ij}$'s. Their effects are easy to see from the mass formula \eqref{mass4140detail}. It is clear that the uncertainties of the coupling parameters $C_{ij}$ also affect the estimated values. We move on to this issue. Because of the complicated quark couplings in tetraquark structures, the properties of two-quark interactions become unclear. To reflect the effects induced by small variations in the coupling parameters, a dimensionless constant
\begin{eqnarray}\label{Kij}
K_{ij}=\frac{\partial E_{CMI}}{\partial C_{ij}},
\end{eqnarray}
can be defined \cite{Li:2018vhp,Liu:2019zoy,Cheng:2020nho}. With this constant, the CMI eigenvalue can be written as
\begin{eqnarray}\label{ECMIwithKij}
E_{CMI}=\sum_{i<j}K_{ij}C_{ij}.
\end{eqnarray}
One should note that this formula \eqref{ECMIwithKij} does not mean that $E_{CMI}$ is the linear superposition of the $C_{ij}$ because the value of $K_{ij}$ also relies on $C_{ij}$. With the $K_{ij}$ amplitudes, one can roughly understand the influence that the coupling parameters $C_{ij}$ have on the estimated tetraquark masses. The $K_{ij}$ values that we calculated are listed in tables~\ref{KijofccQq}-\ref{KijofbcQq}. The results show that the effect on the tetraquark masses due to the uncertainty of $C_{ij}$ depends on the states. For example, the uncertainties of $C_{cc}$, $C_{c\bar{c}}$, $C_{c\bar{n}}$, and $C_{cn}$ have equal effects on the $2^+$ $cc\bar{c}\bar{n}$ state mass, while those of $C_{c\bar{c}}$ and $C_{c\bar{n}}$ have larger effects on the ground $0^+$ $cc\bar{c}\bar{n}$ than $C_{cc}$ and $C_{cn}$ do. As for the tetraquark masses uncertainties, they could even be tens of MeV if those of $C_{c\bar{c}}$ and $C_{c\bar{n}}$ are both 1 MeV. It is also observed that the effects in the $n$ and $s$ cases are not so different.

\begin{table}[htbp]
\caption{K factors of CMI eigenvalues for $cc\bar{c}\bar{n}$, $cc\bar{c}\bar{s}$, $cc\bar{b}\bar{n}$, and $cc\bar{b}\bar{s}$ states.}\label{KijofccQq}
		\begin{tabular}{ccccccc|ccccccc}\hline\hline 
			System&$J^P$&Mass&\multicolumn{4}{c}{K factors}&System&$J^P$&Mass&\multicolumn{4}{c}{K factors}\\\hline
			$cc\bar{c}\bar{n}$&&&$K_{cc}$&$K_{c\bar{c}}$&$K_{c\bar{n}}$&$K_{cn}$& $cc\bar{c}\bar{s}$&&&$K_{cc}$&$K_{c\bar{c}}$&$K_{c\bar{s}}$&$K_{cs}$\\
			&$2^+$&$\left(\begin{array}{c}5372.0\end{array}\right)$&$\begin{array}{c}2.7\end{array}$&$\begin{array}{c}2.7\end{array}$&$\begin{array}{c}2.7\end{array}$&$\begin{array}{c}2.7\end{array}$ & &$2^+$&$\left(\begin{array}{c}5463.7\end{array}\right)$&$\begin{array}{c}2.7\end{array}$&$\begin{array}{c}2.7\end{array}$&$\begin{array}{c}2.7\end{array}$&$\begin{array}{c}2.7\end{array}$\\
			&$1^+$&$\left(\begin{array}{c}5382.5\\5309.8\\5242.4\end{array}\right)$&$\begin{array}{c}3.5\\2.7\\3.2\end{array}$&$\begin{array}{c}4.9\\-4.6\\-3.0\end{array}$&$\begin{array}{c}6.0\\-0.8\\-7.9\end{array}$&$\begin{array}{c}-3.9\\2.4\\-5.2\end{array}$&&$1^+$&$\left(\begin{array}{c}5472.5\\5401.0\\5330.6\end{array}\right)$&$\begin{array}{c}3.5\\2.7\\3.2\end{array}$&$\begin{array}{c}4.8\\-4.5\\-3.0\end{array}$&$\begin{array}{c}6.0\\-0.8\\-7.9\end{array}$&$\begin{array}{c}-3.8\\2.3\\-5.2\end{array}$\\
			&$0^+$&$\left(\begin{array}{c}5435.5\\5191.1\end{array}\right)$&$\begin{array}{c}3.5\\3.1\end{array}$&$\begin{array}{c}7.5\\-12.8\end{array}$&$\begin{array}{c}7.5\\-12.8\end{array}$&$\begin{array}{c}3.5\\3.1\end{array}$&&$0^+$&$\left(\begin{array}{c}5527.9\\5281.4\end{array}\right)$&$\begin{array}{c}3.5\\3.1\end{array}$&$\begin{array}{c}7.5\\-12.8\end{array}$&$\begin{array}{c}7.5\\-12.8\end{array}$&$\begin{array}{c}3.5\\3.1\end{array}$\\
			\hline
			
			$cc\bar{b}\bar{n}$&&&$K_{cc}$&$K_{b\bar{c}}$&$K_{c\bar{n}}$&$K_{bn}$ & $cc\bar{b}\bar{s}$&&&$K_{cc}$&$K_{b\bar{c}}$&$K_{c\bar{s}}$&$K_{bs}$\\
			&$2^+$&$\left(\begin{array}{c}8699.1\end{array}\right)$&$\begin{array}{c}2.7\end{array}$&$\begin{array}{c}2.7\end{array}$&$\begin{array}{c}2.7\end{array}$&$\begin{array}{c}2.7\end{array}$ & &$2^+$&$\left(\begin{array}{c}8789.9\end{array}\right)$&$\begin{array}{c}2.7\end{array}$&$\begin{array}{c}2.7\end{array}$&$\begin{array}{c}2.7\end{array}$&$\begin{array}{c}2.7\end{array}$
			\\
			
			&$1^+$&$\left(\begin{array}{c}8723.5\\8658.7\\8594.6\end{array}\right)$&$\begin{array}{c}3.4\\2.7\\3.2\end{array}$&$\begin{array}{c}4.2\\-8.7\\1.9\end{array}$&$\begin{array}{c}6.5\\2.7\\-11.9\end{array}$&$\begin{array}{c}-3.9\\0.3\\-3.1\end{array}$
			&&$1^+$&$\left(\begin{array}{c}8814.7\\8749.5\\8684.0\end{array}\right)$&$\begin{array}{c}3.4\\2.7\\3.2\end{array}$&$\begin{array}{c}4.1\\-8.7\\1.9\end{array}$&$\begin{array}{c}6.5\\2.7\\-11.9\end{array}$&$\begin{array}{c}-3.9\\0.3\\-3.1\end{array}$\\
			
			&$0^+$&$\left(\begin{array}{c}8750.6\\8547.9\end{array}\right)$&$\begin{array}{c}3.5\\3.1\end{array}$&$\begin{array}{c}7.5\\-12.8\end{array}$&$\begin{array}{c}7.5\\-12.8\end{array}$&$\begin{array}{c}3.5\\3.1\end{array}$
			&&$0^+$&$\left(\begin{array}{c}8842.0\\8637.2\end{array}\right)$&$\begin{array}{c}3.5\\3.1\end{array}$&$\begin{array}{c}7.5\\-12.8\end{array}$&$\begin{array}{c}7.5\\-12.8\end{array}$&$\begin{array}{c}3.5\\3.1\end{array}$\\
			\hline\hline
		\end{tabular}
\end{table}

\begin{table}[htbp]
	\caption{K factors of CMI eigenvalues for $bb\bar{c}\bar{n}$, $bb\bar{c}\bar{s}$, $bb\bar{b}\bar{n}$, and $bb\bar{b}\bar{s}$ states.}\label{KijofbbQq}
		\begin{tabular}{ccccccc|ccccccc}\hline\hline 
			System&$J^P$&Mass&\multicolumn{4}{c}{K factors}& System&$J^P$&Mass&\multicolumn{4}{c}{K factors}\\\hline
			$bb\bar{c}\bar{n}$&&&$K_{bb}$&$K_{b\bar{c}}$&$K_{b\bar{n}}$&$K_{cn}$& $bb\bar{c}\bar{s}$&&&$K_{bb}$&$K_{b\bar{c}}$&$K_{b\bar{s}}$&$K_{cs}$\\
			
			&$2^+$&$\left(\begin{array}{c}12020.9\end{array}\right)$&$\begin{array}{c}2.7\end{array}$&$\begin{array}{c}2.7\end{array}$&$\begin{array}{c}2.7\end{array}$&$\begin{array}{c}2.7\end{array}$& &$2^+$&$\left(\begin{array}{c}12122.9\end{array}\right)$&$\begin{array}{c}2.7\end{array}$&$\begin{array}{c}2.7\end{array}$&$\begin{array}{c}2.7\end{array}$&$\begin{array}{c}2.7\end{array}$\\
			
			&$1^+$&$\left(\begin{array}{c}12013.9\\11992.2\\11942.9\end{array}\right)$&$\begin{array}{c}3.6\\2.7\\3.1\end{array}$&$\begin{array}{c}7.0\\-1.9\\-7.8\end{array}$&$\begin{array}{c}1.8\\-2.0\\-2.4\end{array}$&$\begin{array}{c}-2.6\\1.7\\-5.7\end{array}$ &
			&$1^+$&$\left(\begin{array}{c}12114.2\\12093.0\\12041.3\end{array}\right)$&$\begin{array}{c}3.6\\2.7\\3.0\end{array}$&$\begin{array}{c}6.8\\-2.2\\-7.2\end{array}$&$\begin{array}{c}2.3\\-2.0\\-3.0\end{array}$&$\begin{array}{c}-2.7\\2.0\\-5.9\end{array}$\\
			
			&$0^+$&$\left(\begin{array}{c}12052.0\\11940.0\end{array}\right)$&$\begin{array}{c}3.6\\3.1\end{array}$&$\begin{array}{c}7.5\\-12.8\end{array}$&$\begin{array}{c}7.5\\-12.8\end{array}$&$\begin{array}{c}3.6\\3.1\end{array}$ &
			&$0^+$&$\left(\begin{array}{c}12155.3\\12039.1\end{array}\right)$&$\begin{array}{c}3.6\\3.1\end{array}$&$\begin{array}{c}7.5\\-12.8\end{array}$&$\begin{array}{c}7.5\\-12.8\end{array}$&$\begin{array}{c}3.6\\3.1\end{array}$\\
			\hline

			$bb\bar{b}\bar{n}$&&&$K_{bb}$&$K_{b\bar{b}}$&$K_{b\bar{n}}$&$K_{bn}$ & $bb\bar{b}\bar{s}$&&&$K_{bb}$&$K_{b\bar{b}}$&$K_{b\bar{s}}$&$K_{bs}$\\
			
			&$2^+$&$\left(\begin{array}{c}15352.8\end{array}\right)$&$\begin{array}{c}2.7\end{array}$&$\begin{array}{c}2.7\end{array}$&$\begin{array}{c}2.7\end{array}$&$\begin{array}{c}2.7\end{array}$ & &$2^+$&$\left(\begin{array}{c}15454.1\end{array}\right)$&$\begin{array}{c}2.7\end{array}$&$\begin{array}{c}2.7\end{array}$&$\begin{array}{c}2.7\end{array}$&$\begin{array}{c}2.7\end{array}$\\
			
			&$1^+$&$\left(\begin{array}{c}15360.3\\15327.5\\15300.8\end{array}\right)$&$\begin{array}{c}3.5\\2.7\\3.2\end{array}$&$\begin{array}{c}6.2\\0.4\\-9.2\end{array}$&$\begin{array}{c}4.8\\-5.9\\-1.6\end{array}$&$\begin{array}{c}-3.9\\2.0\\-4.8\end{array}$ & &$1^+$&$\left(\begin{array}{c}15462.0\\15427.1\\15401.1\end{array}\right)$&$\begin{array}{c}3.5\\2.7\\3.2\end{array}$&$\begin{array}{c}6.0\\-0.2\\-8.4\end{array}$&$\begin{array}{c}5.0\\-5.2\\-2.5\end{array}$&$\begin{array}{c}-4.0\\2.3\\-5.0\end{array}$\\
			
			&$0^+$&$\left(\begin{array}{c}15379.7\\15276.9\end{array}\right)$&$\begin{array}{c}3.5\\3.1\end{array}$&$\begin{array}{c}7.5\\-12.8\end{array}$&$\begin{array}{c}7.5\\-12.8\end{array}$&$\begin{array}{c}3.5\\3.1\end{array}$ & &$0^+$&$\left(\begin{array}{c}15481.9\\15375.1\end{array}\right)$&$\begin{array}{c}3.5\\3.1\end{array}$&$\begin{array}{c}7.5\\-12.8\end{array}$&$\begin{array}{c}7.5\\-12.8\end{array}$&$\begin{array}{c}3.5\\3.1\end{array}$
			\\\hline\hline
		\end{tabular}
\end{table}

\begin{table}[h!]
	\caption{K factors of CMI eigenvalues for $bc\bar{c}\bar{n}$, $bc\bar{c}\bar{s}$, $bc\bar{b}\bar{n}$, and $bc\bar{b}\bar{s}$ states.}\label{KijofbcQq}
		\begin{tabular}{ccccccccc}
			\hline\hline System&$J^P$&Mass&\multicolumn{6}{c}{K factors}\\\hline$bc\bar{c}\bar{n}$&&&$K_{bc}$&$K_{b\bar{c}}$&$K_{b\bar{n}}$&$K_{c\bar{c}}$&$K_{c\bar{n}}$&$K_{cn}$\\&$2^+$&$\left(\begin{array}{c}8713.9\\8696.2\end{array}\right)$&$\begin{array}{c}-0.5\\1.9\end{array}$&$\begin{array}{c}5.2\\-0.5\end{array}$&$\begin{array}{c}0.7\\4.0\end{array}$&$\begin{array}{c}0.7\\4.0\end{array}$&$\begin{array}{c}5.2\\0.5\end{array}$&$\begin{array}{c}-0.5\\1.9\end{array}$\\&$1^+$&$\left(\begin{array}{c}8719.4\\8698.4\\8665.3\\8633.9\\8602.3\\8555.9\end{array}\right)$&$\begin{array}{c}-2.6\\2.7\\0.1\\1.0\\-3.3\\0.8\end{array}$&$\begin{array}{c}3.8\\0.3\\-1.8\\-10.5\\-0.2\\3.8\end{array}$&$\begin{array}{c}-0.5\\4.2\\-6.0\\-6.4\\1.7\\2.3\end{array}$&$\begin{array}{c}2.1\\4.2\\1.6\\2.4\\-7.9\\-7.0\end{array}$&$\begin{array}{c}4.6\\1.7\\0.7\\3.4\\-3.8\\-11.3\end{array}$&$\begin{array}{c}2.8\\-2.7\\2.5\\-3.9\\3.0\\-3.1\end{array}$\\&$0^+$&$\left(\begin{array}{c}8746.8\\8668.0\\8572.2\\8478.1\end{array}\right)$&$\begin{array}{c}3.5\\-5.9\\2.7\\-3.0\end{array}$&$\begin{array}{c}4.0\\2.6\\-3.7\\-12.2\end{array}$&$\begin{array}{c}3.5\\2.3\\-9.2\\-5.9\end{array}$&$\begin{array}{c}3.5\\2.3\\-9.2\\-5.9\end{array}$&$\begin{array}{c}4.0\\2.6\\-3.7\\-12.2\end{array}$&$\begin{array}{c}3.5\\-5.9\\2.7\\-3.0\end{array}$\\\hline$bc\bar{c}\bar{s}$&&&$K_{bc}$&$K_{b\bar{c}}$&$K_{b\bar{s}}$&$K_{c\bar{c}}$&$K_{c\bar{s}}$&$K_{cs}$\\&$2^+$&$\left(\begin{array}{c}8805.0\\8788.1\end{array}\right)$&$\begin{array}{c}-0.5\\1.9\end{array}$&$\begin{array}{c}5.2\\-0.5\end{array}$&$\begin{array}{c}0.6\\4.0\end{array}$&$\begin{array}{c}0.6\\4.0\end{array}$&$\begin{array}{c}5.2\\-0.5\end{array}$&$\begin{array}{c}-0.5\\1.9\end{array}$\\&$1^+$&$\left(\begin{array}{c}8811.3\\8789.2\\8755.6\\8722.4\\8693.8\\8644.8\end{array}\right)$&$\begin{array}{c}-2.8\\2.9\\0.2\\0.9\\-3.3\\0.8\end{array}$&$\begin{array}{c}3.6\\0.4\\-2.4\\-10.1\\-0.1\\3.9\end{array}$&$\begin{array}{c}-0.1\\4.0\\-5.4\\-7.0\\1.7\\2.1\end{array}$&$\begin{array}{c}2.2\\4.0\\1.4\\2.5\\-8.0\\-6.7\end{array}$&$\begin{array}{c}4.5\\1.9\\0.8\\3.4\\-3.6\\-11.5\end{array}$&$\begin{array}{c}2.9\\-2.9\\2.6\\-3.9\\3.0\\-3.1\end{array}$\\&$0^+$&$\left(\begin{array}{c}8839.6\\8757.5\\8661.5\\8565.4\end{array}\right)$&$\begin{array}{c}3.5\\-5.9\\2.8\\-3.0\end{array}$&$\begin{array}{c}3.9\\2.6\\-3.9\\-12.0\end{array}$&$\begin{array}{c}3.5\\2.3\\-9.0\\-6.1\end{array}$&$\begin{array}{c}3.5\\2.3\\-9.0\\-6.1\end{array}$&$\begin{array}{c}3.9\\2.6\\-3.9\\-12.0\end{array}$&$\begin{array}{c}3.5\\-5.9\\2.8\\-3.0\end{array}$\\\hline$bc\bar{b}\bar{n}$&&&$K_{bc}$&$K_{b\bar{b}}$&$K_{b\bar{n}}$&$K_{b\bar{c}}$&$K_{c\bar{n}}$&$K_{bn}$\\&$2^+$&$\left(\begin{array}{c}12042.0\\12013.5\end{array}\right)$&$\begin{array}{c}-0.5\\1.8\end{array}$&$\begin{array}{c}5.2\\-0.5\end{array}$&$\begin{array}{c}0.6\\4.1\end{array}$&$\begin{array}{c}0.6\\4.1\end{array}$&$\begin{array}{c}5.2\\-0.5\end{array}$&$\begin{array}{c}-0.5\\1.8\end{array}$\\&$1^+$&$\left(\begin{array}{c}12042.1\\12023.1\\11987.3\\11973.2\\11947.5\\11897.5\end{array}\right)$&$\begin{array}{c}0.7\\-0.7\\0.9\\1.2\\-3.0\\-0.4\end{array}$&$\begin{array}{c}5.3\\-1.1\\-4.0\\-8.2\\-1.8\\5.2\end{array}$&$\begin{array}{c}-0.2\\4.7\\-2.3\\-9.5\\2.9\\-0.3\end{array}$&$\begin{array}{c}0.6\\5.1\\0.0\\3.0\\-12.1\\-1.2\end{array}$&$\begin{array}{c}5.2\\1.4\\1.3\\2.8\\0.0\\-15.4\end{array}$&$\begin{array}{c}-1.0\\0.5\\2.3\\-3.5\\1.3\\-1.0\end{array}$\\&$0^+$&$\left(\begin{array}{c}12059.3\\12008.3\\11918.1\\11829.5\end{array}\right)$&$\begin{array}{c}3.3\\-5.8\\1.8\\-2.0\end{array}$&$\begin{array}{c}4.3\\2.5\\-1.6\\-14.5\end{array}$&$\begin{array}{c}3.2\\2.4\\-11.8\\-3.2\end{array}$&$\begin{array}{c}3.2\\2.4\\-11.8\\-3.2\end{array}$&$\begin{array}{c}4.3\\2.5\\-1.6\\-14.5\end{array}$&$\begin{array}{c}3.3\\-5.8\\1.8\\-2.0\end{array}$\\\hline$bc\bar{b}\bar{s}$&&&$K_{bc}$&$K_{b\bar{b}}$&$K_{b\bar{s}}$&$K_{b\bar{c}}$&$K_{c\bar{s}}$&$K_{bs}$\\&$2^+$&$\left(\begin{array}{c}12143.3\\12114.9\end{array}\right)$&$\begin{array}{c}-0.5\\1.9\end{array}$&$\begin{array}{c}5.2\\-0.5\end{array}$&$\begin{array}{c}0.7\\4.0\end{array}$&$\begin{array}{c}0.7\\4.0\end{array}$&$\begin{array}{c}5.2\\-0.5\end{array}$&$\begin{array}{c}-0.5\\1.9\end{array}$\\&$1^+$&$\left(\begin{array}{c}12143.3\\12124.9\\12087.7\\12072.2\\12048.7\\11996.6\end{array}\right)$&$\begin{array}{c}0.7\\-0.8\\1.0\\1.1\\-3.0\\-0.5\end{array}$&$\begin{array}{c}5.3\\-1.1\\-5.1\\-7.3\\-1.7\\5.2\end{array}$&$\begin{array}{c}0.0\\4.8\\-1.5\\-10.0\\2.6\\-0.4\end{array}$&$\begin{array}{c}0.5\\5.1\\-0.3\\2.9\\-11.8\\-1.1\end{array}$&$\begin{array}{c}5.2\\1.4\\1.6\\2.5\\0.0\\-15.4\end{array}$&$\begin{array}{c}-1.1\\0.6\\2.1\\-3.1\\1.1\\-0.9\end{array}$\\&$0^+$&$\left(\begin{array}{c}12161.0\\12109.7\\12016.3\\11928.2\end{array}\right)$&$\begin{array}{c}3.3\\-5.9\\1.9\\-2.1\end{array}$&$\begin{array}{c}4.2\\2.5\\-1.7\\-14.4\end{array}$&$\begin{array}{c}3.2\\2.5\\-11.7\\-3.3\end{array}$&$\begin{array}{c}3.2\\2.5\\-11.7\\-3.3\end{array}$&$\begin{array}{c}4.2\\2.5\\-1.7\\-14.4\end{array}$&$\begin{array}{c}3.3\\-5.9\\1.9\\-2.1\end{array}$\\\hline
		\end{tabular}
\end{table}

Our mass predictions rely on the $X(4140)$ state as a $1^{++}$ reference tetraquark. In previous theoretical works \cite{Wu:2016gas,Stancu:2009ka,Lu:2016cwr,Lebed:2016yvr,Anwar:2018sol}, $X(4140)$ was regarded as a $1^{++}$ $cs\bar{c}\bar{s}$ tetraquark. However, this exotic state's inner structure has not been confirmed in experiments and it is still a subject of theoretical debate. If this state is something other than a tetraquark, one has to consider the effects of this reference assumption on the predictions. For example, a detailed study of $cs\bar{c}\bar{s}$ states in the chiral quark model \cite{Yang:2019dxd} did not obtain a $cs\bar{c}\bar{s}$ tetraquark that was consistent with the observed $X(4140)$, while the authors of Ref. \cite{Bokade:2024tge} interpreted $X(4140)$ as the charmonium state $\chi_{c1}(3P)$ according to their calculation in a relativistic screened potential model. In this situation, all the tetraquark states' masses would change some of the values in our framework since the mass splittings between the $QQ\bar{Q}\bar{q}$ tetraquark states remain unaffected. What we need from the adopted assumption is actually the determination of a tetraquark mass (input scale of the approach). If the observed $X(4140)$ is a mixed structure of charmonium, a molecule, and a compact $cs\bar{c}\bar{s}$ tetraquark, one anticipates that the theoretical mass of the compact $cs\bar{c}\bar{s}$ would not be far from the $D_s^{*+}D_s^{*-}$ threshold; otherwise, the mixing would not be significant. As a result, the shifted value would not be very large. This theoretical scale's determination depends on the proportion of the compact $cs\bar{c}\bar{s}$ in the wave function of $X(4140)$. If $X(4140)$ does not contain a $cs\bar{c}\bar{s}$ component, one may determine the shifted value by treating another compact tetraquark candidate as a reference. In Ref. \cite{Yang:2019dxd}, the interpretation of $X(4274)$ and $X(4350)$ as ground $1^{++}$ and $0^{++}$ $cs\bar{c}\bar{s}$ tetraquarks, respectively, was proposed. If $X(4274)$ is indeed the ground $1^{++}$ $cs\bar{c}\bar{s}$ and we take its mass as the input scale, our predictions for all the tetraquark masses would shift upward by about $M_{X(4274)}-M_{X(4140)}\approx 140$ MeV.

For exotic hadrons, one has to confirm whether they exist through experimental measurements. Most states that we considered here have the quantum numbers of $D^{(*)}$, $D_s^{(*)}$, $\bar{B}^{(*)}$, or $\bar{B}_s^{(*)}$, but with much higher masses. Such resonances can be searched for in the invariant mass distributions of a heavy quarkonium and a $Q\bar{q}$ meson in high-energy colliders, such as the LHC and future CEPC. The $cc\bar{b}\bar{q}$ and $bb\bar{c}\bar{q}$ states are explicitly exotic and they can be searched for similarly.

Here, we considered only a simple rearrangement scheme in which the decay appears to occur through quark components free collisions. In principle, the decay parameter ${\cal C}$ may be evaluated with the quark-level wave functions of the initial and final states. The gluon exchange contributions certainly affect the ${\cal C}$ value further. Such a contribution can probably be explored in a similar way to the quark interchange model in Ref. \cite{Barnes:2000hu}. Because the ${\cal C}$ parameter varies with the state, it is possible that its variation may significantly alter the predicted decay width ratios. However, the spatial wave functions should be obtained for such a consideration. If one wants to include gluon exchange contributions to ${\cal C}$ but without explicit spatial wave functions, additional parameters would be introduced, which are not easy to determine with the available experimental data.

Replacing the light antiquark in a triply heavy tetraquark state with a light diquark (triquark) would produce a triply heavy pentaquark (hexaquark) state. The present framework can be extended to study such systems. In the extension, one would confront the problem of how to select appropriate reference scales that are consistent with tetraquark studies. We will consider this problem in future investigations.

To summarize, we studied the properties of triply heavy tetraquark states in this work. We estimated their spectra in a modified CMI model by treating $X(4140)$ as a reference $1^{++}$ tetraquark. No stable state was found. We also considered their two-body strong decays and the related indicative partial width ratios of different channels in a simple rearrangement scheme. We hope that our results can help future experimental searches for such exotic states.

\section{ACKNOWLEDGMENTS}

This project was supported by the National Natural Science Foundation of China (Nos. 12235008, 12275157, 12475143, 11905114) and the Shandong Province Natural Science Foundation (ZR2023MA041).

\end{document}